\documentclass[11pt,graphicx,amsmath]{article}
\usepackage{amsmath}
\usepackage{graphicx}
\usepackage{bm}
\usepackage[dvips]{color}
\usepackage{amssymb}
\usepackage{amsfonts}
\usepackage{comment}
\usepackage{cite}
\usepackage{todonotes}
\usepackage{caption}
\usepackage{subcaption}

\def\be{\begin{equation}}
\def\ee{\end{equation}}
\def\nn{\nonumber}

\newcommand{\red}[1]{{\color{red} #1}}
\def\ba{\begin{eqnarray}}
\def\ea{\end{eqnarray}}
\def\bl#1\el{\begin{align}#1\end{align}}

\def\l{\left}
\def\r{\right}

\title{ A massless scalar field in  Robertson-Walker  spacetimes:
        Adiabatic regularization  and Green's  function
                   }
\author{\small
            \,  Yang  Zhang\thanks{yzh@ustc.edu.cn} , \,
            Bo Wang   \thanks{ymwangbo@ustc.edu.cn} \,
            and Xuan Ye   \thanks{yyyyy@mail.ustc.edu.cn} \\
 \small  Department of  Astronomy,
         CAS Key Laboratory for Researches in Galaxies and Cosmology, \\
 \small  School of Astronomy and Space Sciences, \\
 \small  University of Science and Technology of China, Hefei, Anhui, 230026, China \\
 }

 \date{}

\def\be{\begin{equation}}
\def\ee{\end{equation}}
\def\nn{\nonumber}

\large

\begin{document}

\maketitle

\begin{abstract}

We study  adiabatic regularization of
a coupling  massless scalar field
in general spatially flat  Robertson-Walker (RW)  spacetimes.
For the conformally-coupling,
the 0th-order regularized power spectrum and 0th-order
regularized stress tensor are zero,
and  no trace anomaly exists in general RW spacetimes.
This is a new result which extents those found in de Sitter space.
For the minimally-coupling,
the regularized spectra are also zero
in the radiation-dominant stage, the matter-dominant stage,
and   de Sitter space as well.
The vanishing of these adiabatically regularized spectra are also confirmed
by direct regularization of the Green's functions.
For a general coupling and general RW spacetimes,
the regularized spectra can
be negative under the conventional prescription.
By going to higher order of regularization,
the spectra will generally become  positive,
but will also acquire IR divergence which is inevitable for a massless field.
To avoid the IR divergence,
the inside-horizon regularization is applied.
By these procedures,    one will eventually achieve
nonnegative,  UV- and IR-convergent power spectrum and spectral energy density.

\end{abstract}

\

{\bf quantum fields in curved spacetimes},
{\bf inflationary universe},
{\bf mathematical and relativistic  aspects of cosmology},


\large

\section{Introduction}

The vacuum expectation values of stress tensor and power spectrum of
quantum fields in curved spacetimes have direct observational effects
in cosmology.
However, these physical quantities are prone to  UV divergences
 \cite{UtiyamaDeWitt1962,DeWitt1975,FeynmanHibbs1965}.
To remove UV divergences,
several approaches have been proposed for regularization,
such as the dimensional regularization
\cite{CandelasRaine1975,DowkerCritchley1976,Brown1977,Bunch1979},
the covariant point-splitting \cite{Christensen1976,AdlerLiebermanNg1977,DaviesFullingChristensenBunch1977,
Christensen1978,BunchDavies1978,BunchChristensenFulling1978,Wald1978},
and the zeta function \cite{DowkerCritchley1976,Hawking1977,Fujikawa1980}.
These methods involve the Green's function
and are essentially  equivalent \cite{DowkerCritchley1976}.
The appropriate subtraction term to  the Green's function
in position space is generally hard to determine.
Only for the massless scalar field in de Sitter space
with conformal, or  minimal coupling,
the   subtraction term  has been found   \cite{ZhangYeWang2019}.
Different from the above approaches,
the adiabatic regularization works with the $k$-modes
\cite{ParkerFulling1974,FullingParkerHu1974,
HuParker1978,BLHu1978,Birrell1978,
Bunch1978,Bunch1980,AndersonParker1987,BunchParker1979,
BirrellDavies1982,ParkerToms,Parker2007,Markkanen2018,
WangZhangChen2016,ZhangWangJCAP2018},
and by the minimal subtraction rule
the power spectrum is regularized to the 2nd-order,
and the stress tensor to the 4th-order.
For a massive scalar field
with $\omega=(k^2+m^2)^{1/2}$  as the 0th-order frequency,
this prescription is sufficient in removing all UV divergences,
but sometimes removes more than necessary,
and  leads to negative spectra,
 as demonstrated in de Sitter space \cite{ZhangYeWang2019}.
In fact,
 0th-order  regularization is sufficient to achieve
nonnegative, UV- and IR-convergent spectra
for a massive scalar field with  conformal coupling,
and, similarly, so is   2nd-order regularization  for  minimal coupling.
Given the regularized power spectrum,
Fourier transformation produces
the regularized Green's function
which is UV and IR convergent  \cite{ZhangYeWang2019}.

In this paper we extend the study to   general spatially flat RW spacetimes.
We shall consider a  massless scalar field with a coupling $\xi$,
whose exact solution is available
and regularization can be performed in an analytical manner.
Our aim is to search for proper regularization schemes
which will yield nonnegative,  UV- and IR-convergent
power spectrum  and spectral energy density.
As shall be shown, the goal can be eventually achieved,
but  there is no universal scheme
that would work for all couplings and all RW spacetimes.
First, UV convergence can be easily achieved by  regularization of certain order.
Unlike a massive field,
the 0th-order frequency of a massless field is the wavenumber $k$,
so that  the 2nd-order regularization for the power spectrum
and  the 4th-order for the spectral stress tensor
are   necessary to remove all UV divergences.
In particular, for the conformal coupling $\xi=\frac16$
the  regularization of all orders are equivalent,
  yielding   a zero power spectrum and a zero spectral stress tensor,
and there is no trace anomaly.
For the minimal coupling $\xi=0$,
the regularized power spectrum and  spectral  stress tensor
are zero  in several important RW spacetimes,
such as  the radiation-dominated (RD) expansion,
the  matter-dominated  (MD) expansion,
and the de Sitter space.
For general couplings and general RW spacetimes,   however,
the regularized spectra can be  negative, as demonstrated herein.
In order to avoid this  negative spectrum,
we attempt   to increase the order of regularization on the pertinent spectrum
as this  will generally yield  a positive spectrum.
Nevertheless, this higher-order regularized, positive  spectrum
tends to    carry new IR divergence
which  is characteristic of
the adiabatically  regularized spectra of a massless field \cite{ZhangWangJCAP2018}.
To retain     IR convergence,
we shall   apply the inside-horizon scheme of regularization,
by which the long wavelength modes outside the horizon are fixed
and only the short wavelength modes inside the horizon
are regularized  \cite{ZhangWangJCAP2018}.
Finally  we shall  achieve
nonnegative, UV- and IR-convergent
power spectrum  and spectral energy density.

The paper is organized as follows.

In Sec. \ref{section2}, we derive the exact solution of a coupling massless scalar field
and its Green's function in   general RW spacetimes,
analyze the behaviors of the power spectrum and the spectral stress tensor,
and give the prescriptions of adiabatic regularization.

In Sect. \ref{section3},   for the conformal coupling $\xi=\frac16$,
we show that  adiabatic regularization of various orders are equal,
yielding zero power spectrum and zero stress tensor,
and there is no trace anomaly in   general RW spacetimes.
We also give direct regularization of the  Green's function,
 confirming the result of adiabatic regularization.

Sect. \ref{section4}  considers   the minimal  coupling   $\xi=0$.
We show that the regularized power spectrum and stress tensor are  zero
in several important  RW spacetimes,
and the results are also confirmed by regularization of the Green's functions.
In   general RW  spacetimes,
we use  two examples to show that  regularized spectra can be negative,
and the pertinent spectrum will become positive
by realizing  to  higher-order regularization,
 thereby also acquire IR divergence.
The IR divergence  will be avoided by the inside-horizon  scheme.

Sect. \ref{section5} presents the case   for the general coupling  $\xi$,
the analysis of which is similar to  Sect. \ref{section4}.

Sect. \ref{section6}  provides   conclusion  and  discussions.

Appendix \ref{sectionA}   lists  high-$k$ expansions of the exact modes.
Appendix \ref{sectionB}   lists  the WKB solutions
      and the associated subtraction terms  up to 6th-order,
      and demonstrates  the covariant conservation to each adiabatic order.

\section{ The massless scalar field in general RW spacetimes}\label{section2}

In  a flat Robertson-Walker spacetime
\be \label{metric}
ds^2=a^2(\tau)[d\tau^2- \delta_{ij}   dx^idx^j],
\ee
with the conformal time $\tau$,
a massless scalar field has
the Lagrangian density
\be
{\cal L} =\frac12 \sqrt{-g}
 \l( g^{\mu\nu}\phi_{,\mu}\phi_{,\nu} -\xi R\phi^2 \r) ,
\ee
and the  field  equation   \cite{DeWitt1975,BunchDavies1978,Bunch1980}
\be    \label{fieldequxi}
\l(  \Box  + \xi R  \r)\phi =0 ,
\ee
where $\Box = \frac{1}{a^4} \frac{\partial}{\partial \tau}
(a^2 \frac{\partial}{\partial \tau}) -\frac{1}{a^2} \nabla^2$,
$R= 6 a''/a^{3}$ is  the scalar curvature,
and $\xi $ is the   coupling constant,
and we consider a range  $0 \leq \xi \leq \frac16$  specifically;
and
\[
\phi  ({\bf x},\tau) = \int\frac{d^3k}{(2\pi)^{3/2}}
        \left[ a _{\bf k}   \phi_k(\tau)  e^{i\bf{k}\cdot\bf{x}}
    +a^{\dagger}_{\bf k} \phi^{*}_k(\tau) e^{-i\bf{k}\cdot\bf{x}}\right]
\]
where  $a_{\bf k}, a^{\dagger}_{\bf k} $ are
the annihilation and creation operators, respectively,
that satisfy the   canonical commutation relation,
and $\phi_k(\tau)$ is the $k$-mode,  written as
$\phi_k(\tau) =  v_k(\tau)/a(\tau)$.
The  equation of the  rescaled $v_k$ mode is
\be\label{equvk}
v_k'' + \Big( k^2   + (\xi  -\frac16 ) a^2 R  \Big) v_k = 0 .
\ee
In this paper
we  consider a class of power-law expanding  RW spacetimes,
\be \label{inflation}
a(\tau)= a_0 |\tau|^{b} ,
\ee
where the expansion index $b$ is a constant,
$a^2 R = 6 b(b-1)  \tau^{-2 }$.
The   positive-frequency mode solution of (\ref{equvk}) is
\be  \label{u}
v_k (\tau )  \equiv  \sqrt{\frac{\pi}{2}}\sqrt{\frac{x}{2k}}
  e^{i \frac{\pi}{2}(\nu+ \frac12) } H^{(1)}_{\nu} ( x) ,
\ee
where    $x \equiv  k |\tau|$,
$ H^{(1)}_{\nu}$ is the Hankel functions, and
\be\label{nubrelation}
\nu \equiv  \sqrt{\frac14    -( 6\xi   -1)  b(b-1)} .
\ee
Three cases lead to a special value  $\nu=\frac12$:
the conformal coupling ($\xi=\frac16$), the Minkowski spacetime  ($b=0$),
and the RD expansion  ($b=1$),
and in these cases
 the  mode (\ref{u}) reduces to
\be  \label{u016}
v_k (\tau ) =   i \sqrt{\frac{\pi}{2}}\sqrt{\frac{x}{2k}}
  H^{(1)}_{\frac12} ( x)
  =    \frac{1}{\sqrt{2k} }   e^{-i k\tau} ,
\ee
conformal to the mode in  Minkowski spacetime.
The mode (\ref{u}) of a general $\nu$ at high $k$
 also approaches to (\ref{u016}).

\begin{figure}[htb]
\centering
        \includegraphics[width = .6\linewidth]{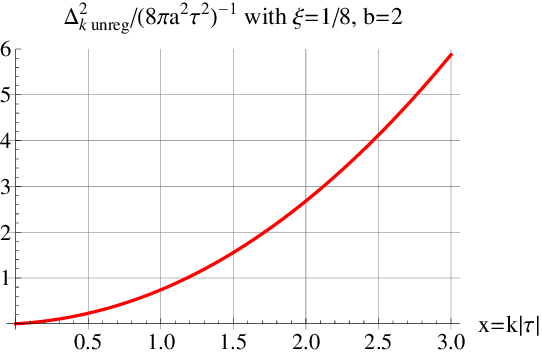}
\caption{
     The unregularized $\Delta^2_k $  for  $b=2$ and   $\xi=\frac18$.
     The plot is with $|\tau|=1$ for illustration.
    }
     \label{2b12xipowerSpect}
\end{figure}

The Bunch-Davies vacuum state
 is defined   such that
\be\label{ask}
a_{\bf k}  |0 \rangle =0, ~~~
      {\rm for\   all} \  {\bf k} .
\ee
In a general RW spacetime
the  unregularized  Green's function   in the vacuum  is
\ba \label{GreenInteg}
G(x^\mu, x'\, ^\mu) &
    = &  \langle0|  \phi(\textbf{r},\tau) \phi (\textbf{r}',\tau') |0\rangle
        =   \frac{1}{(2\pi)^3} \int d^3k \, e^{i \bf k \cdot (r-r')}
           \phi_k(\tau) \phi_{k}^* (\tau ')  \nn \\
& = & \frac{1}{ a(\tau)a(\tau')}
  \frac{|\tau|^{1/2} |\tau'|^{1/2} }{8 \pi }
    \int_0^\infty  dk  k    \frac{\sin( k|r-r'|)}{ |r-r'|  }
  H^{(1)}_{\nu} (k\tau ) H^{(2)}_{\nu} (k\tau') .
\ea
The  integration (\ref{GreenInteg})
can be carried out \cite{BunchDavies1978,Watson1958,ZhangYeWang2019},
and the result is a hypergeometric function
as  follows
\be\label{GreeHyper}
G(\sigma)=   \frac{1}{16 \pi^2  a(\tau)a(\tau') |\tau \tau'| }
\Gamma \big( \frac{3}{2}-\nu \big) \Gamma \big( \nu +\frac{3}{2} \big)
 \, _2 F_1  \left[\frac{3}{2}+\nu ,\frac{3}{2}-\nu ,2, ~  1 + \frac{\sigma}{2}
    \right] \, ,
\ee
where  $\sigma=  [(\tau-\tau')^2-(r-r')^2 ]/(2 \tau \tau')$
is the half of squared geometric
distance between two points $x^\mu$ and $x'^\mu$.
For the equal-time $\tau=\tau'$ case the Green's function   is
\ba\label{Greenunregdf}
G({\bf r}- {\bf r}')
   =  \langle0|  \phi(\textbf{r},\tau) \phi (\textbf{r}',\tau) |0\rangle
   =  \int_0^\infty  \frac{\sin( k|r-r'|)}{|r-r'|\, k^2}
          \Delta^2_k (\tau)  \, d k ,
\ea
 the auto-correlation function  is
\bl\label{vevcorr}
G(0)= \langle0|  \phi(\textbf{r},\tau) \phi (\textbf{r},\tau) |0\rangle
         =   \frac{1}{(2\pi)^3} \int d^3k \,  |\phi_k(\tau)|^2
         = \int_0^{\infty}\Delta_k^2 (\tau)\frac{dk}{k}  ,
\el
and the   power spectrum   is
\be \label{BunchDaviesSpectrum}
\Delta_k^2 (\tau)
=\frac{ k^{3}}{2  \pi^2 a^2 (\tau) }   |v_k(\tau)|^2
=\frac{  x^{3} }{8  \pi a^2(\tau)\tau^{2} } | H^{(1)}_{\nu} ( x)|^2 .
\ee
The   power spectrum is shown in Fig.\ref{2b12xipowerSpect}
for  $b=2$ and   $\xi=\frac18$,
which is   UV divergent,  leading  to an infinite  auto-correlation $G(0)$.
This also indicates that the stress  tensor in (\ref{Tmunu})
as well as  the trace in (\ref{traceTmunu})  will be generally divergent
as they  contains a term $\propto \phi^2$.
The asymptotic  behaviors of power spectrum can be analyzed by the series expansion.
At high-$k$,   the power spectrum for general $\xi$ and $b$ is
\bl\label{psex}
\Delta_k^2 (\tau)
=\frac{ 1}{4  \pi^2 a^2\tau^2 }\Big[
 &
 x^{2}-\frac{(6 \xi -1) (b-1) b}{2 }
      -\frac{3(6 \xi -1) (b-1) b \l[(1-6\xi ) (b^2-b)-2\r]}{8 x^2} \nn \\
&    -\frac{5(6 \xi -1)(b-1) b\l[(1-6\xi ) (b^2-b)-2\r]
   \l[ (1-6\xi ) (b^2-b)-6\r] }{16 x^4}       + ...  \Big] .
\el
The first two terms are quadratic and  logarithmic
UV divergent.
Our task is to perform adiabatic regularization
and establish  a power spectrum which  should  be:
1)    UV convergent,
2) IR convergent,
3) nonnegative.
By the minimal subtraction rule \cite{Parker2007},
the first two  terms of (\ref{psex}) are  subtracted off
under the 2nd-order regularization,
\be \label{regps2nd}
\Delta^2_{k\, reg} = \frac{ k^{3}}{2  \pi^2 a^2 }
      \Big( |v_k(\tau)|^2 -|v_k^{(2)}(\tau)|^2 \Big),
\ee
where $|v_k^{(2)}(\tau)|^2$ is the subtraction term
given by  (\ref{v2subsq}),
formed from the 2nd-order WKB  approximate solution.
The 2nd-order regularized power spectrum (\ref{regps2nd})
is  UV convergent;
however, it can be  negative for certain values of $\xi$ and $b$,
as can be checked by the  dominant, third term of Eq.(\ref{psex}),
\be \label{3thtermx}
 -\frac{3(6 \xi -1) (b-1) b \l[(1-6\xi ) (b^2-b)-2\r]}{8 x^2}
\ee
at high $k$.
When this term is negative,   to obtain a positive power spectrum,
we shall try  the 4th-order  regularized   power spectrum,
\be \label{regps4th}
\Delta^2_{k\, reg} = \frac{ k^{3}}{2  \pi^2 a^2 }
      \Big( |v_k(\tau)|^2 -|v_k^{(4)}(\tau)|^2 \Big),
\ee
where $|v_k^{(4)}(\tau)|^2$ is given by (\ref{v2sub}),
constructed from the 4th-order WKB  approximate solution,
and removes  all the first three terms of Eq.(\ref{psex}).
This usually  yields a positive, UV-convergent power spectrum,
which is dominated at high $k$
by  the fourth term  of Eq.(\ref{psex})
\be\label{4thtermps}
 -\frac{5(6 \xi -1)(b-1) b\l[(1-6\xi ) (b^2-b)-2\r]
   \l[ (1-6\xi ) (b^2-b)-6\r] }{16 x^4} .
\ee
We have  checked that (\ref{4thtermps}) is positive
when  (\ref{3thtermx}) is negative.
We shall demonstrate  this procedure  by  examples in later sections.

We examine the low $k$ behavior of $\Delta^2_{k}$.
For $\xi=\frac{1}{6}$, or $b=1$, it has only one term,
\be
\Delta^2_{k}   =  \frac{ k^{3}}{2  \pi^2 a^2}  \frac{1}{2 k} ,
\ee
which holds also  for all $k$.
For a general $\xi$ by   (\ref{sqvkm}), it is
\be\label{spLowk1}
\Delta_k^2 (\tau)
\simeq \frac{2^{2\nu}}{8 \pi^3 a^2 \tau^{2}}
 \Gamma(\nu)^2  x^{3-2\nu } \propto k^{3-2\nu }.
\ee
Regarding  inflationary cosmology,
if the scalar field $\phi$ is
used to model the perturbed  inflaton scalar field during inflation,
the power spectrum $\Delta_k^2$ at low $k$ realizes
the primordial spectrum of scalar field perturbations,
which   is often written in a form of
\[
\Delta_k^2 \propto k^{\, n_s -1} .
\]
Thus, one reads off the scalar spectral index
\be
n_s= 4-2\nu =4- 2 \sqrt{\frac14    -( 6\xi   -1)  b(b-1)} .
\ee
The currently observed value  is $n_s\simeq 0.96$,
which for $\xi=0$ corresponds to an expansion index $b\simeq -1.02$
during inflation \cite{WangZhangChen2016,ZhangWangJCAP2018}.
In general RW spacetimes,
the power spectrum (\ref{spLowk1}) with $\xi=0$
is IR convergent  for $-1<b<2$,
and  is  IR divergent for  $b \leq -1$  or  $b \geq 2$.
In this paper we do not discuss the issue of IR divergence
in  the unregularized power spectrum,
which can be avoided either by  certain initial condition
or by some precedent expansion stage,
as been studied in Ref. \cite{FordParker1977,VilenkinFord1982}.
Nevertheless,  for a massless field,
sometimes regularization may take an IR convergent power spectrum
into   IR divergent.
 This is one of the things that we are concerned with in this paper.
When this happens,   to retain the IR convergence,
we can adopt the scheme of inside-horizon  regularization,
i.e,
the long wavelength modes outside the horizon are fixed
and only the short wavelength modes inside the horizon
are regularized \cite{ZhangWangJCAP2018}.
By this  procedure,
the regularized power spectrum will remain IR convergent.

The stress tensor plays  the role of a source of gravity in general relativity.
For the  massless scalar field,
it is given by  \cite{Bunch1980,AndersonParker1987}
\bl \label{Tmunu}
T_{\mu\nu} =&  (1-2\xi) \partial_ \mu \phi \partial_ \nu \phi
  +(2\xi - \frac12) g_{\mu\nu } \partial^\sigma  \phi \partial_\sigma  \phi
  -2\xi \phi_{;\mu\nu} \phi
  \nn \\
&  + \frac12 \xi g_{\mu\nu} \phi \Box \phi
 -\xi (R_{\mu\nu}-\frac12 g_{\mu\nu} R + \frac32 \xi R g_{\mu\nu}) \phi^2,
\el
which satisfies the covariant conservation $T^{\mu\nu}_{~~~ ;\nu}   = 0$
by virtue of  the field equation (\ref{fieldequxi}).
The trace is
\be\label{traceTmunu}
T^\mu\, _\mu = (6\xi -1) \partial^ \mu \phi \partial_ \mu \phi
          +\xi (1-6\xi) R \phi^2 .
\ee
The energy density in the BD vacuum state is given by the expectation value
\be \label{energyspectr}
\rho  = \langle T^0\, _0 \rangle =\int^{\infty}_0   \rho_k \frac{d k}{k} ,
\ee
where the spectral energy density is
\bl \label{rhok}
\rho_k
=& \frac{ k^3}{4\pi^2 a^4}
 \Big[ |v_k'|^2 + k^2  |v_k|^2
  + (6\xi-1) \Big(  \frac{a'}{a} (v'_k v^*_k + v_k v^*\, '_k  )
    - (\frac{a'}{a})^2 |v_k|^2  \Big) \Big] ,
\el
the trace of stress tensor   is
\bl
\langle T^\mu\, _\mu \rangle
  =&   \frac{1}{2\pi^2 a^4} \int  k^2 dk \,
  (6\xi-1)\Big [ |v_k'|^2 - \frac{a'}{a} (v'_k v^*_k + v_k v^*\, '_k  )
   - k^2 |v_k|^2   \nn \\
&  -\big(\frac{a''}{a} - (\frac{a'}{a})^2 \big) |v_k|^2
    + (1- 6\xi) \frac{a''}{a} |v_k|^2 \Big]  ,
\el
the pressure  is
\bl
p =   -\frac13 \langle T^i \, _i \rangle
     =  \int^\infty_0   p_k \frac{dk}{k} ,
\el
where  the spectral pressure is
\bl \label{sprectpressure}
p_k =&  \frac{k^3}{4 \pi^2 a^4}
  \Big[   \frac13 |v_k'|^2 + \frac13 k^2  |v_k|^2
  + 2(\xi-\frac16)\Big(  \frac{a'}{a} (v'_k v^*_k + v_k v_k^*\, ' )
    - (\frac{a'}{a})^2 |v_k|^2 \Big)    \nn \\
& - 4(\xi-\frac16)\Big( |v_k'|^2 - \frac{a'}{a} (v'_k v^*_k + v_k v_k^*\, ')
  - k^2  |v_k|^2
   - (\frac{a''}{a} - (\frac{a'}{a})^2) |v_k|^2 \nn \\
&   - 6(\xi-\frac16) \frac{a''}{a} |v_k|^2  \Big)\Big]  .
\el
For each $k$-mode,
the  unregularized spectral stress tensor  satisfies the covariant conservation
$\rho_{k }' +  3 \frac{a'}{a} (\rho_{k} +  p_{k}) =0$,
as can be checked by the field  equation (\ref{equvk}).
The spectral energy density and pressure have
  the following  high-$k$ expansions
\bl \label{rhokHighExp}
\rho_k
=\frac{1}{4\pi^2 a^4\tau^4}\Big[
&
x^4 -\frac{(6 \xi-1)b^2 x^2}{2 }+\frac{3 (6 \xi -1)^2(b-1)b^2 (b+1) }{8}
\nn\\
&
+\frac{5(6 \xi -1)^2 (b-1) b^2 (b+2)\l[(1- 6\xi)  (b^2-b)-2\r] }{16 x^2}
 \nn \\
&  +\frac{35 (6 \xi -1)^2b^2 (b-1) (b+3) [b (b-1) (6 \xi -1)+2 ]
   [(b-1) b (6 \xi -1)+6 ]}{128 x^4}
\nn \\
&    + ...\Big] ,
\el
\bl \label{presskHighExp}
p_k = \frac{1}{4 \pi^2 a^4\tau^4} \Big[
& \frac{x^4}{3}
-\frac{(6 \xi -1)b (b+2) x^2}{6} +\frac{ (6 \xi -1)^2(b-1)b(b+1) (b+4)}{8}
\nn\\
& + \frac{5(6 \xi -1)^2(b-1) b(b+2) (b+6)  \l[(1- 6\xi)  (b^2-b)-2\r]}{48 x^2} \nn \\
& + \frac{35\text{  }(6 \xi -1)^2 (b-1) b (b+3) (b+8) [ (b-1) b (6 \xi -1)+2]
   [(b-1) b (6 \xi -1)+6]}{384 x^4} \nn \\
& +... \Big]   ,
\el
both containing   quartic,   quadratic,  and  logarithmic UV divergences.
 The quartic divergences come from the derivative terms,
such as $\partial_ \mu \phi \partial_ \nu \phi$,
other than the $\phi^2$ term.
We search for a regularized stress tensor
that should satisfy  the following criteria :
1)   UV convergent,
2)   IR  convergent,
3)  the spectral energy density is nonnegative.
A negative pressure   is allowed,
so we focus on the spectral energy density in this paper.
By the minimal subtraction rule \cite{ParkerFulling1974},
the first three divergent terms of $\rho_k$ in (\ref{rhokHighExp})
are  subtracted off under  the 4th-order regularization,
\be\label{enerreg4}
 \rho_{k\, reg} =  \rho_{k} - \rho_{k\, A 4 }  ,
\ee
where $\rho_{k\, A 4}$ is  the 4th-order subtraction term
given by (\ref{rhoA42countx}).
The resulting $\rho_{k\, reg}$ is always  UV convergent,
and if it is also positive, our goal is achieved.
Like  the   power spectrum,
however,  the 4th-order regularized $\rho_k$   can be negative
for certain values of $\xi$ and  $b$,
as is determined the dominant fourth term of Eq.(\ref{rhokHighExp}) at high $k$,
\be \label{c4thter}
\frac{5(6 \xi -1)^2 (b-1) b^2 (b+2) \Big((1- 6\xi)  (b^2-b)-2 \Big) }{16 x^2} .
\ee
When this happens,
in order to get a positive spectral energy density,
we shall try the 6th-order regularization
\be \label{c6thter}
 \rho_{k\, reg} =  \rho_{k} - \rho_{k\, A 6 }  ,
\ee
where the subtraction term $\rho_{k\, A 6}$
is given by Eq.(\ref{rhoA6countx}).
Under this subtraction,
the first four terms of (\ref{rhokHighExp})  will be removed,
and   the fifth term
\be \label{c6thtermenergy}
\frac{35 (6 \xi -1)^2b^2 (b-1) (b+3) \Big(b (b-1) (6 \xi -1)+2 \Big)
   \Big((b-1) b (6 \xi -1)+6 \Big) }{128 x^4}
\ee
remains and dominates the 6th-order regularized spectral energy density at high $k$.
As we shall see, for  RW spacetimes used  in cosmology,
this often  gives a positive,  UV-convergent
regularized spectral energy density.
For   RW spacetimes  rarely used in cosmology, say $-3<b<-2 $,
and for  a coupling
\be \label{rangeNeg}
 \frac{b^2-b-2}{6 b^2-6 b}<\xi <\frac{1}{6},
\ee
both Eq.(\ref{c4thter}) and Eq.(\ref{c6thtermenergy}) will be negative for high $k$.
Then one has to seek the 8th-order regularized $\rho_k$
which  is dominated by the  following term
\bl
\frac{1}{4\pi^2 a^4\tau^4}\Big[
&
-\frac{1}{256 x^6}63 (b-1) (b+4) b^2(1-6 \xi )^2
\Big((b-1) b (6 \xi -1)
+2\Big) \Big((b-1) b (6 \xi -1) \nn\\
&
+6\Big) \Big((b-1) b (6 \xi -1)+12\Big)
\Big] .
\el
This term is positive with a coupling in the range of (\ref{rangeNeg}).
Thus, a positive UV-convergent spectral energy density
 can be achieved by this procedure.

We  mention that
the four-divergence of the subtraction terms to
the stress tensor is zero under regularization of each order,
so that  the covariant conservation is respected
by the  regularized stress tensor respects
(See (\ref{TA0})--(\ref{cconservres})  in Appendix \ref{sectionB}.)

\begin{figure}[htb]
\centering
        \includegraphics[width = .6\linewidth]{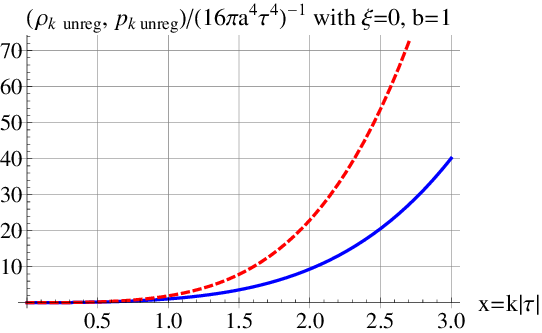}
\caption{
Red Dash: unregularized $\rho_k $,
    Blue Solid: unregularized  $p_k $.
    For  $\xi=0$ and $b=1$.
    }
     \label{energyUN01}
\end{figure}

We examine the low-$k$  behavior of the stress tensor.
For  $\xi= \frac16$, or   $b=1$,
Eq.(\ref{nubrelation}) gives  $\nu=\frac12$,
and (\ref{rhok}) (\ref{sprectpressure}) reduce to
\be \label{rhok16}
\rho_k= \frac{ k^3}{4\pi^2 a^4}
\Big[ |v_k'|^2 + k^2  |v_k|^2     \Big]
= \frac{ k^3}{4\pi^2 a^4} k   ,
~~~~
p_k=  \frac13 \rho_k   ,
\ee
which are  UV divergent and IR convergent.
For    $\xi=0$,  (\ref{rhok}) (\ref{sprectpressure}) reduce to
\bl\label{energykxi0}
\rho_k  =&  \frac{ k^3}{4\pi^2 a^2}
 \Big[  |(\frac{v_k}{a})' |^2  + k^2  |\frac{v_k}{a} |^2  \Big] ,
\el
\bl  \label{pkxi0}
p_k =&   \frac{k^3}{4 \pi^2 a^2}
  \Big[ |(\frac{v_k}{a})' |^2  - \frac13 k^2 |\frac{v_k}{a}|^2  \Big] ,
\el
shown  in Fig. \ref{energyUN01},
and at low $k$  they reduce to
\be  \label{enersmallk}
\rho_k  \simeq  \left\{
\begin{aligned}
    &
 \frac{ 2^{2+2 \nu }}{16\pi^3 a^4\tau^{4}}
\Gamma (\nu +1)^2 x^{3-2 \nu }
    ,
    &b\geq\frac{1}{2},
    \\
    &
    \frac{2^{2-2\nu }}{16\pi a^4\tau^{4}}
\frac{ x^{2 \nu +3}}{\sin^2(\pi  \nu )\Gamma (\nu )^2},
    &0< b<\frac{1}{2},
      \\
   &
   \frac{x^4}{4 \pi ^2 a^4\tau^{4}}     ,
    & b=0,
      \\
   &
   \frac{2^{2\nu}}{16\pi^3 a^4\tau^{4}}
\Gamma(\nu)^2 x^{5-2\nu }  ,  &
   b<0,
\end{aligned}
\right.
\ee
\be\label{presssmallk}
p_k  \simeq  \left\{
\begin{aligned}
    &
 \frac{ 2^{2+2 \nu } }{16\pi^3 a^4\tau^{4}}
\Gamma (\nu +1)^2 x^{3-2 \nu }
    ,
    &b\geq\frac{1}{2},
    \\
    &
    \frac{2^{2-2\nu }}{16\pi a^4\tau^{4}}
\frac{ x^{2 \nu +3}}{\sin^2(\pi  \nu )\Gamma (\nu )^2},
    &0< b<\frac{1}{2},
      \\
   & \frac{ x^4}{12 \pi ^2 a^4\tau^{4}},  &
   b=0 ,
      \\
   & - \frac{2^{2\nu}}{48\pi^3 a^4\tau^{4}}
\Gamma(\nu)^2 x^{5-2\nu } ,  &
   b<0 .
\end{aligned}
\right.
\ee
where $\nu=b-\frac{1}{2}$ if $b\geq\frac{1}{2}$,
and $\nu=\frac{1}{2}-b$ if
$b<\frac{1}{2}$.
Both  $\rho_k$ and $p_k$ are IR convergent when $-2<b<2$,
as shown in Fig. \ref{energyUN01},
which includes the RD stage ($b=1$) and the de Sitter inflation ($b=-1$).
But when $b= \pm 2$, or  $b>2$,   or  $b<-2$,
 $\rho_k$ and $p_k$ are  IR divergent.
The issue of IR divergence of the unregularized spectra
has been analyzed in Ref. \cite{FordParker1977,VilenkinFord1982},
and  in this paper we shall not discuss it further.
Regularization may take   IR convergent  $\rho_k$ and $p_k$
into  IR divergent, just as with the  power spectrum,
and  the scheme of inside-horizon regularization
can be used to retain IR convergence \cite{ZhangYeWang2019}.

\section{  Regularization for  $\xi=\frac16$}\label{section3}

First we implement  adiabatic  regularization  of
the conformally-coupling  massless scalar field,
which is an  interesting  case.
This is because   Eq.(\ref{equvk}) with  $\xi=\frac16$
is that same as  that in  the Minkowski spacetime,
ie,
the equation of mode $\phi_k =  v_k /a$
is conformal to that in  the Minkowski spacetime.
Thus, the conformally-coupling  massless scalar field $\phi$
is said to have the conformal symmetry.
Moreover, this conformal symmetry is also reflected
by the zero trace of stress tensor
as presented in  Eq.(\ref{traceTmunu}) with $\xi=\frac16$.
Now  the rescaled mode for $\xi=\frac16$
is given by
\be\label{vkxi16}
v_k(\tau)=\frac{ e^{-i k\tau }}{\sqrt{2 k}},
\ee
which is  valid for any expansion index $b$.
(See  Eqs. (\ref{nubrelation}) and (\ref{u016}).)
The corresponding power spectrum  (\ref{BunchDaviesSpectrum})
has only one term,
\be \label{deltxi16}
\Delta_k^2 (\tau)
=\frac{ k^{3}}{2  \pi^2 a^2 }  \frac{1}{2 k} ,
\ee
so that the  0th-order regularization is   sufficient,
yielding  a  vanishing regularized spectrum
\bl\label{ps16m0}
\Delta^2_{k\, reg} =& \frac{ k^{3}}{2  \pi^2 a^2 }
      \Big( |v_k(\tau)|^2 -|v_k^{(0)}(\tau)|^2 \Big)
      =  \frac{ k^{3}}{2  \pi^2 a^2 }
      \Big( \frac{1}{2 k} -\frac{1}{2 k} \Big) =0.
\el
The  spectral energy density and pressure  in Eq.(\ref{rhok16})
have only one term,
and the trace   $\langle T^\mu\, _\mu \rangle_k=0$.
The 0th-order subtraction terms  are  given  by (\ref{rhoA0}) (\ref{pkA0}),
and are  just equal to the unregularized stress tensor.
Hence,  the regularized stress tensor is zero,
\be \label{Regenergy0}
\rho_{k\, reg}
= \rho_{k} -\rho_{k\, A 0} = \frac{ k^4}{4\pi^2 a^4} -\frac{ k^4}{4\pi^2 a^4} =0 ,
\ee
\be \label{Regpress0}
p_{k\, reg} =  p_{k} -p_{k\, A 0}
=  \frac{ k^4}{12\pi^2 a^4} - \frac{ k^4}{12\pi^2 a^4}=0 ,
\ee
and  the regularized trace is also zero,
\be\label{zerostress}
  \langle T^{\beta}\, _\beta \rangle _{k\, reg}=
  \rho_{k\, reg} -3 p_{k\, reg} =  0 .
\ee
The above calculations   show  two important  features of
the  conformally-coupling massless scalar field.
First,   the vanishing spectra
(\ref{ps16m0})--(\ref{zerostress}) hold  for a general scale factor $a(\tau)$.
Thus,  in any flat RW spacetime,
the power spectrum and the stress tensor are all regularized to zero
and  there is no trace anomaly
 of the conformally-coupling massless scalar field.
This is a generalization
of the result in de Sitter space \cite{ZhangYeWang2019}
to general RW spacetimes.
Second, for  $\xi=\frac16$,
the vanishing regularized spectra (\ref{ps16m0})--(\ref{zerostress})
hold for any order of adiabatic regularization,
because
the subtraction terms of any order are  equal to those  of the 0th-order.
(See (\ref{W246})--(\ref{pressureA246})  in Appendix \ref{sectionB}).

The above results of adiabatic regularization
also follow from a direct regularization of  Green's function.
For   $\nu=\frac12$,
the unregularized Green's function  (\ref{GreeHyper}) reduces to
\be\label{m0xi16Green}
G(\sigma)= - \frac{1}{16 \pi^2 a(\tau)a(\tau') \tau \tau'}  \frac{2}{\sigma}   ,
\ee
which has one term,   and is UV divergent at $\sigma=0$.
To remove this  UV divergence,
the natural  choice for the  subtraction term   is
\be\label{subGrm0xi16}
G(\sigma)_{sub} = - \frac{1}{16 \pi^2 a(\tau)a(\tau') \tau \tau'}\frac{2}{\sigma} ,
\ee
and
the regularized   Green's  function
is simply given by
\be\label{regGreenm0Xi16}
G(\sigma )_{reg}  =  G(\sigma)  - G(\sigma)_{sub}  =0 .
\ee
This vanishing Green's function
confirms the vanishing power spectrum of (\ref{ps16m0}),
as  they are the Fourier transformation to each other.
Consequently the regularized stress tensor    is also vanishing
when it is constructed from the vanishing Green's function.

Thus under the above two different approaches
we have demonstrated that
the zero trace is still ensured by the proper regularization.
Two references \cite{Wald1978,DaviesFullingChristensenBunch1977}
also worked directly with a massless scalar field,
and claimed that the trace of stress tensor would become nonzero
(the  so-called trace anomaly) after regularization.
In Eq.(3) of  Ref.\cite{Wald1978}
the Green's function $G(\sigma)$
is assumed to have a term  $v\ln \sigma +w$,
which would lead to the trace anomaly.
As our Eq.(\ref{m0xi16Green}) tells,
the Green's function  for $\xi=\frac16$
contains no such a term,
so that the conclusion in Ref.\cite{Wald1978}
on the existence of trace anomaly does not hold
in RW spacetimes.
We have also examined
Ref.\cite{DaviesFullingChristensenBunch1977}
on a massless scalar field in RW spacetimes,
and find that
their calculated  $\langle T_{\mu\nu} \rangle$ of Eq.(5.30)
does not contain the trace anomaly by any combination,
and that their  trace anomaly  Eq.(6.5) was actually put in by hand,
rather than  following from any finite part of Eq.(5.30).

\section{  Regularization for $\xi=0$}\label{section4}

Next  we perform   regularization for  the minimally-coupling $\xi=0$.
As listed  in  Appendix \ref{sectionB},
the   subtraction terms of various orders depend
on the expansion  index $b$ through $a(\tau)$.
We shall consider some specific values of $b$.
(The case  $b=-1$ of de Sitter space
was realized  in Ref.\cite{ZhangYeWang2019}.)

We first  consider  the RD expansion stage,
in which the index $b=1$ and the scalar curvature $R=0$.
Eq.(\ref{nubrelation}) gives  $\nu=\frac12$
which holds  for any coupling $\xi$.
The rescaled mode $v_k$  becomes the same as (\ref{vkxi16}),
and the  power spectrum $\Delta_k^2 $
becomes the same as (\ref{deltxi16}).
We use the 2nd-order regularization,
and obtain a zero regularized power spectrum
\be\label{deltb1}
\Delta^2_{k\, reg }
=  \frac{ k^{3}}{2  \pi^2 a^2 }
      \Big( |v_k(\tau)|^2 -|v_k^{(2)}(\tau)|^2 \Big)
=  0 .
\ee
Notice that  $a''=a''' = ... = 0$
for the RD stage with $a\propto \tau$,
so that the 0th-, 2nd- and higher order adiabatic  substraction terms
of the power spectrum are equal,
$|v_k^{(0)}(\tau)|^2=|v_k^{(2)}(\tau)|^2 =...= \frac{1}{2k}$.
(See  (\ref{ModvkWKB0})  (\ref{v2subsq})
 (\ref{invW4s}) (\ref{v2sub})
 (\ref{invW6s}) (\ref{v6sub})
in Appendix \ref{sectionB}.)
In addition,  the regularization of  the corresponding Green's function is the same as
those given by  Eqs.(\ref{m0xi16Green})
 (\ref{subGrm0xi16}) (\ref{regGreenm0Xi16}) in the previous  section.
The spectral energy density (\ref{energykxi0})
and spectral pressure (\ref{pkxi0})   become
\be\label{rhob1}
\rho_k
   =  \frac{1}{4\pi^2 a^4\tau^{4}}
 \Big( x^4
 +  \frac{x^2 }{2 }
  \Big) ,
\ee
\be
p_k    =\frac{1}{4\pi^2 a^4\tau^{4}}
 \Big( \frac{x^4}{3 }
 +  \frac{x^2 }{2 }
  \Big) ,
\ee
which contain  quartic and quadratic UV divergences.
The 2nd-order substraction terms  (\ref{rhoA22countx}) (\ref{pA2counxt}) are
\be \label{subenergy4thRD}
 \rho_{k\, A 2} =
  \frac{ k^4}{4 \pi ^2 a^4 } \left(\frac{1}{2 x^2}+1\right),
\ee
\be \label{subpress4thRD}
 p_{k\, A 2}
 =  \frac{ k^4 }{12 \pi ^2a^4} \left(\frac{3}{2 x^2}+1\right) ,
\ee
so   the regularized stress tensor are vanishing
\be \label{Regenergy4thRD}
\rho_{k\, reg}
= \rho_{k} -\rho_{k\, A 2}    =0 ,
\ee
\be \label{Regpress4thRD}
p_{k\, reg}
=  p_{k} -p_{k\, A 2}  =0 .
\ee
(Note that  the 4th-order substraction terms   (\ref{rhoA42countx}) (\ref{pA4counxt})
happen to be equal to those  of the 2nd-order as  $a''=a'''=0$.)
Thus, in the RD stage, the 2nd-order regularization
is sufficient to remove
all the divergences of the power spectrum and stress tensor
of a minimally-coupling massless field.
This is similar to that which occurs  in de Sitter space,
in which the 2nd-order regularization also produces a zero
the power spectrum and zero stress tensor
of a minimally-coupling  massless scalar field  \cite{ZhangYeWang2019}.
The result can be also  derived in terms of Green's function.
For   $b=1$,
the unregularized Green's function  (\ref{GreeHyper}) is the same as (\ref{m0xi16Green}),
and so is the  subtraction term  in (\ref{subGrm0xi16}),
so the regularized   Green's  function is
$G(\sigma )_{reg}  =  G(\sigma)  - G(\sigma)_{sub}  =0$;
thus  the regularized power spectrum  is zero
and the regularized stress tensor is also zero.

We  next   consider   the matter-dominated (MD) stage, $b=2$.
The mode (\ref{u}) with $\nu=\frac32$ becomes
\be
v_k (\tau )  = \frac{e^{i x} }{\sqrt{2k}} (1+\frac{i}{x}) ,
\ee
and the  power spectrum (\ref{BunchDaviesSpectrum})  becomes
\be\label{psxim0}
\Delta_k^2 (\tau)
=\frac{ k^{3}}{2  \pi^2 a^2 } \l( \frac{1}{2 k}
        +\frac{1}{2 k^3\tau^2}\r) .
\ee
By the  2nd-order  regularization, using $|v_k^{(2)}|^2$  from   (\ref{v2subsq}),
 the regularized spectrum  is  zero
\be \label{PSxi0m0}
\Delta^2_{k\, reg } =  \frac{ k^{3}}{2  \pi^2 a^2 }
      \Big( |v_k(\tau)|^2 -|v_k^{(2)}(\tau)|^2 \Big)
       = 0 \, .
\ee
The spectral stress tensor   (\ref{energykxi0})  (\ref{pkxi0}) for $b=2$ becomes
\be
\rho_k     =  \frac{1}{4\pi^2 a^4\tau^{4}}
 \Big(x^4
 +  2x^2
 +  \frac{9}{2 }
  \Big)  ,
\ee
\be
p_k =   \frac{1}{4\pi^2 a^4\tau^{4}}
 \Big( \frac{x^4}{3}
+ \frac{4x^2}{3}
+\frac{9}{2}
  \Big) ,
\ee
which contain  quartic, quadratic and logarithmic  UV divergences.
To remove these divergences,
we apply  the 4th-order regularization,
and the subtraction terms  (\ref{rhoA42countx}) and (\ref{pA4counxt}) are
\be \label{subRegenergy4th}
\rho_{k\, A 4} =  \frac{ k^4}{4 \pi ^2 a^4 }
\left(1 +\frac{2}{x^2} + \frac{9}{2 x^4} \right) ,
\ee
\be \label{subRegpress4th}
p_{k\, A 4}  = \frac{ k^4}{12 \pi ^2 a^4 }
\left(1 +\frac{4}{x^2} + \frac{27}{2 x^4}   \right) .
\ee
 The  regularized spectral energy density and pressure are zero
\be \label{Regenergy4th}
\rho_{k\, reg} =  \rho_{k} -\rho_{k\, A 4}=0
,
\ee
\be \label{Regpress4th}
p_{k\, reg} =  p_{k} -p_{k\, A 4}=0 .
\ee
These results can be also   derived in terms of Green's function.
For  $\nu=\frac32$ we can directly integrate  Eq.(\ref{GreenInteg})
to obtain the Green's function
\bl
G(\sigma)
&=\frac{1}{(2\pi)^3  } \frac{1}{  a(\tau)a(\tau')}
  \int \frac{1}{k}d^3k \,  e^{i {\bf k\cdot(r-r')}-ik(\tau-\tau')}
   \frac12 \bigg(1+i (\frac{1}{\tau'}-\frac{1}{\tau})\frac{1}{k}
  + \frac{1}{k^2}\frac{1}{\tau\tau' } \bigg) \nn \\
&  = \frac{ 1}{8\pi^2 a(\tau)a(\tau') |\tau\tau'|  }
           \big( - \frac{1}{\sigma} - \ln  \sigma  \big) ,
\el
in which both terms  are UV convergent and should be removed.
So the subtraction term is taken to be
\be
G(\sigma)_{sub} =\frac{ 1}{8\pi^2 a(\tau)a(\tau') |\tau\tau'|}
          \big( - \frac{1}{\sigma} - \ln  \sigma  \big),
\ee
resulting in
\be\label{xi0reps}
G(\sigma)_{reg }= G(\sigma)  - G(\sigma)_{sub} =0 \, ,
\ee
which  agrees with
the  vanishing  regularized   spectra
(\ref{PSxi0m0}) (\ref{Regenergy4th}) and (\ref{Regpress4th}).

From the spectra (\ref{psex}) (\ref{rhokHighExp}) and (\ref{presskHighExp})
with $\xi=0$,
we see that, after subtracting their  respective  divergent terms,
all the remaining convergent terms are proportional
a common  factor
\[
b (b-2) (b-1) (b+1) ,
\]
which is vanishing for $b = 0,\pm 1, 2$.
Hence,  for the minimally-coupling massless scalar field,
the regularized power spectrum
and stress tensor are  zero
in the   Minkowski spacetime,
the RD stage, the MD stage,
and the de Sitter space,
the latter case was shown in Ref. \cite{ZhangYeWang2019}.

\begin{figure}
\centering
\includegraphics[width=0.6\linewidth]{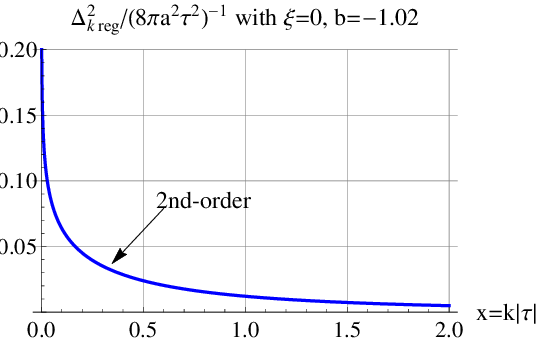}
\caption{
   For  $\xi=0$, $b=-1.02$:
the 2nd-order $\Delta_{k\,reg}^2$ is UV convergent and IR divergent.
      }
     \label{powerReb102Xi0new}
\end{figure}

\begin{figure}
\centering
\includegraphics[width=0.6\linewidth]{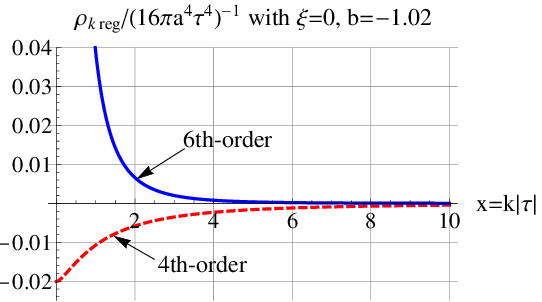}
\caption{
    Red Dash:  the 4th-order regularized $\rho_{k\, reg}$ is negative,
    Blue Solid: the 6th-order regularized $\rho_{k\, reg}$
    is positive,  and  $k^{-4}$ UV convergent  at high $k$.
    The model  $b=-1.02$ and $\xi=0$.
      }
     \label{64thrhob12xi0}
\end{figure}

\begin{figure}
\centering
\includegraphics[width=0.6\linewidth]{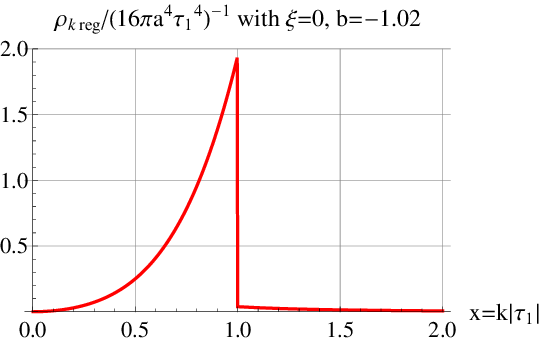}
\caption{
For  $\xi=0$, $b=-1.02$:
the inside-horizon regularization for $\rho_k(\tau)_{reg}$
according to Eq.(\ref{Hrhoreg}).
     The plot is   at a time $|\tau_1|=1$ for illustration.
      }
     \label{rhoInsideH}
\end{figure}

What about a  general index  $b$?
In the following we consider two  quasi de Sitter inflation models with $b\simeq  -1$.
For the model  $b=-1.02$,
as shown in Fig. \ref{powerReb102Xi0new},
the 2nd-order regularized  $\Delta_{k\,reg}^2$ is  positive, UV convergent and IR divergent.
However,   the 4th-order regularized  $\rho_{k\,reg}$ is negative
 as shown in Fig. \ref{64thrhob12xi0}.
This is implied by
the dominant,   fourth term of (\ref{rhokHighExp}) of $\rho_k$,
\be  \label{4threm}
\frac{5 (b-2) (b-1) b^2 (b+1) (b+2)}{16 x^2} ,
\ee
which  is negative for $b=-1.02$.
This is the phenomenon of negative spectra that we have mentioned early
around  Eqs.(\ref{c4thter}) (\ref{c6thter}).
To obtain  a positive spectral energy density,
we  proceed to compute the 6th-order regularized $\rho_{k\,reg}$ in Eq.(\ref{c6thter}),
which is dominated by the  fifth  term of (\ref{rhokHighExp}).
As a result,  the 6th-order regularized $\rho_{k\,reg}$
 is positive and  UV convergent, as shown in Fig.\ref{64thrhob12xi0}.
As for the  IR divergences in the regularized spectra,
we adopt the inside-horizon scheme  \cite{ZhangWangJCAP2018}
as  follows.
The UV divergences come from the high $k$ modes;
whereas,  the low  $k$ modes do not cause  UV divergence.
Therefore,   only   the short wavelength modes need to be regularized
inside the horizon during the expansion ($k \gtrsim 1/|\tau_1|$,
where  $\tau_1$ is a fixed time during the expansion),
and the long wavelength modes outside the  horizon remain  unchanged.
Under this scheme,  UV divergences are removed and IR divergences are avoided.
Thus, for the case  $b=-1.02$  under  consideration,
  the spectral energy density is regularized by
\ba \label{Hrhoreg}
 \rho_k(\tau)_{reg}  =
\Bigg \{
 \begin{array}{ccc}
      \rho_{k} - \rho_{k\, A 6 }  ,
  \,\,  & \text{ for $k \ge \frac{1}{|\tau_1|}$} ,
    \\
     \rho_{k} ,
    \,\,\,\,\,  & \text{ for $k < \frac{1}{|\tau_1|}$} \red{.}
   \\
 \end{array}
\ea
The regularization is performed  instantaneously at a fixed $\tau_1$.
The result is  plotted in Fig. \ref{rhoInsideH}.
Thus,  the IR divergence is avoided,
and a positive,  UV- and IR-convergent spectral energy density is  achieved,
and the regularized energy density  is
$\rho_{reg}=  \int_0^\infty  \rho_{k\, reg}   \frac{dk}{k}
    =0.622  \frac{(b/a_0)^4}{16\pi}$  at $|\tau|=1$.

\begin{figure}
\centering
\includegraphics[width=0.6\linewidth]{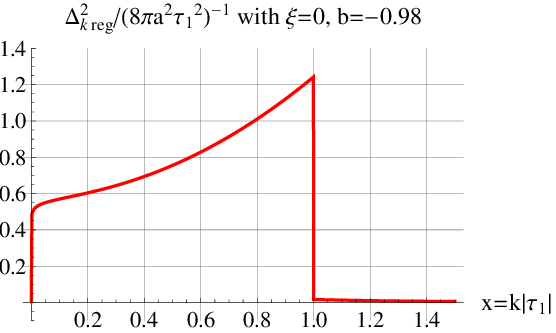}
\caption{
   For  $\xi=0$, $b=-0.98$:
the inside-horizon regularization for $\Delta^2_{k\, reg}$
according to Eq.(\ref{regvac2counth2}).
       }
     \label{powerInsideH}
\end{figure}

\begin{figure}
\centering
\includegraphics[width=0.6\linewidth]{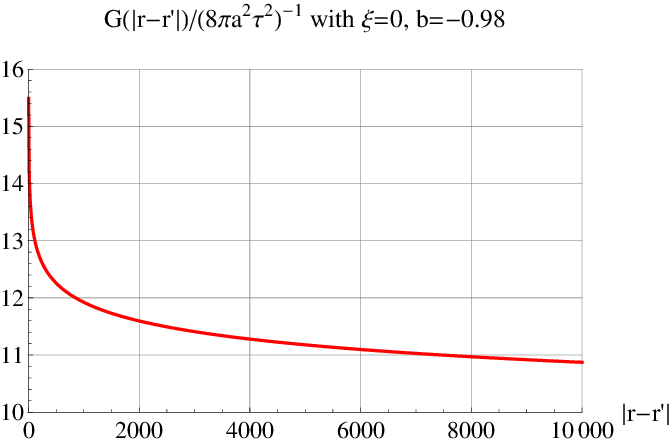}
\caption{
  For  $\xi=0$, $b=-0.98$:
the regularized Green's function $G(|{\bf r}- {\bf r}'|)_{reg}$,
is Fourier transform of the regularized  power spectrum in Fig.\ref{powerInsideH}.
         }
     \label{GreenInsideH}
\end{figure}

\begin{figure}
\centering
\includegraphics[width=0.6\linewidth]{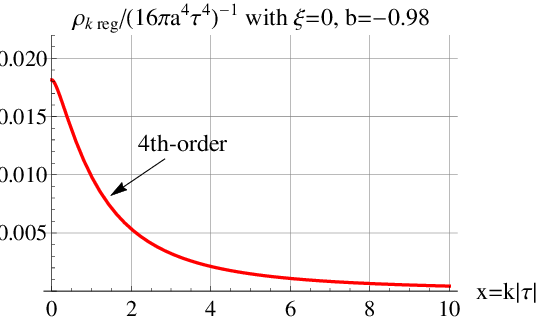}
\caption{
   For  $\xi=0, b=-0.98$:
   the 4th-order $\rho_{k\,reg}$ is positive,
     UV convergent and IR log divergent.     }
     \label{4thrReb98Xi0new}
\end{figure}

For the model $b=-0.98$,
\begin{figure}
\centering
\includegraphics[width=0.6\linewidth]{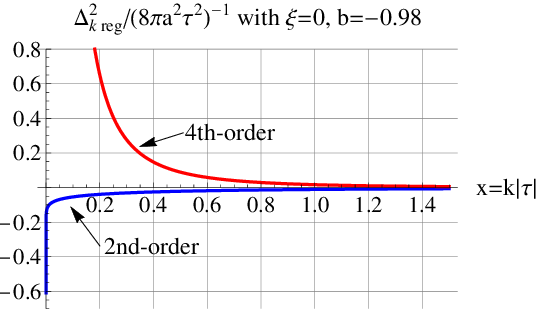}
\caption{
   For  $\xi=0, b=-0.98$:
the 2nd-order $\Delta_{k\,reg}^2$ is negative;
the 4th-order $\Delta_{k\,reg}^2$ is UV convergent and IR divergent.
       }
     \label{powerReb98Xi0new}
\end{figure}
the 2nd-order regularized power spectrum is dominated by the third term of (\ref{psex})
and is negative,  as shown in Fig. \ref{powerReb98Xi0new}.
Thus,   we proceed to calculate
the 4th-order regularized power spectrum according to Eq.(\ref{regps4th}),
which  is positive, UV-convergent,
dominated by the fourth term of (\ref{psex}), as shown in Fig. \ref{powerReb98Xi0new}.
To avoid the IR divergence  caused by the 4th-order regularization,
we apply the inside-horizon scheme
\be \label{regvac2counth2}
\Delta_k^2( \tau)_{reg} =\frac{ k^3}{2 \pi^2 a^2}
\l \{
 \begin{array}{lcc}
    \big( |v_{k}|^2 - |v_k^{(4)}|^2 \big)   ,
  \,\,  & \text{ for $k \ge \frac{1}{|\tau_1|}$} ,
    \\
   |v_{k}|^2,
    \,\,\,\,\,  & \text{ for $k  <  \frac{1}{|\tau_1|}$} .
   \\
 \end{array}
\r.
\ee
The resulting power spectrum  is  plotted in Fig. \ref{powerInsideH}.
Accordingly,   the IR divergence is avoided,
and a positive,  UV- and IR-convergent power spectrum  is  achieved.
By the Fourier transformation of (\ref{regvac2counth2})
according to the formula (\ref{Greenunregdf}),
we obtain the corresponding regularized Green's function
$G({\bf r}- {\bf r}')_{reg}$, which  is UV finite and IR convergent,
as shown in Fig. \ref{GreenInsideH}.
The 4th-order regularized spectral energy density
 is positive and UV convergent,
as  plotted in Fig. \ref{4thrReb98Xi0new},
and the regularized energy density  is
$\rho_{reg} =  \int_0^\infty  \rho_{k\, reg}   \frac{dk}{k}
    =0.228\frac{(b/a_0)^4}{16\pi}$  at $|\tau|=1$.
The above  two examples show that
the inside-horizon scheme is effective in avoiding
IR divergences.

\section{  Regularization for general $\xi$}\label{section5}

Now we explore  adiabatic regularization
for   a general coupling $\xi$,
search for proper regularization schemes
that would yield nonnegative,  UV- and IR-convergent
spectra $\Delta^2_{k }$  and $\rho_k$.
We shall consider several values of  $b$
in several  interesting cosmological models.
By the minimal subtraction rule,
the 2nd-order regularization  for  $\Delta^2_{k }$
and the 4th-order regularization  for $\rho_k$  are default,
and we shall attempt  higher order regularization
 when a negative spectrum  appears.

First we consider
 $b=1$ for the RD stage with a general $\xi$.
The analysis between (\ref{deltb1})--(\ref{Regpress4thRD})
is also valid for  a general $\xi$,
and the results are
$\Delta^2_{k\, reg }=\rho_{k\, reg }=p_{k\, reg}=0$.

\begin{figure}
\centering
\includegraphics[width=0.6\linewidth]{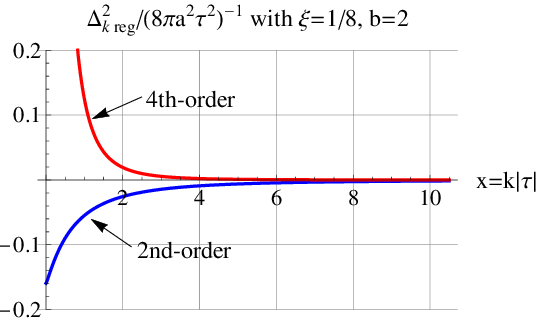}
\caption{
 For $b=2$ and $\xi=  1/8 $,
the 2nd-order   regularized $\Delta_k^2$ is negative,
and the 4th-order  regularized $\Delta_k^2$ is positive, and UV convergent.
      }
     \label{powerReb2Xi18new}
\end{figure}

\begin{figure}
\centering
\includegraphics[width=0.6\linewidth]{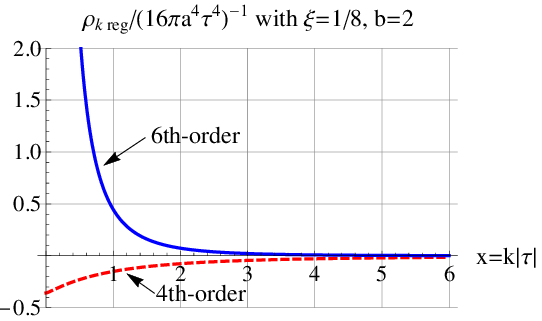}
\caption{
  For $b=2$ and $\xi=  \frac18  $,
  Red Dashed: the 4th-order regularized $\rho_{k\, reg}$ is negative;
  Blue Solid:  the 6th-order regularized $\rho_{k\, reg}$
    is positive,  and  $k^{-4}$ UV convergent at high $k$.
      }
     \label{64thrhob2xi8new}
\end{figure}

Next we consider $b=2$ for the MD stage with a general $\xi$.
For  illustration,   $\xi= \frac{1}{8}$ is taken in the following,
other values of $\xi$ can be analyzed in the same fashion.
As shown in Fig. \ref{powerReb2Xi18new},
the 2nd-order regularized $\Delta_{k\, reg}^2$ is  negative,
dominated by the   third term of Eq.(\ref{psex}),
and the 4th-order regularized $\Delta_{k\, reg}^2$
 is positive and UV convergent, but IR divergent.
As shown in Fig.\ref{64thrhob2xi8new},
the 4th-order regularized $\rho_{k\, reg}$  is negative,
dominated by the fourth term   in  Eq.(\ref{rhokHighExp}),
and the 6th-order regularized  $\rho_{k\, reg}$  is positive,
dominated by the  fifth  term of (\ref{rhokHighExp}).

\begin{figure}
\centering
\includegraphics[width=0.6\linewidth]{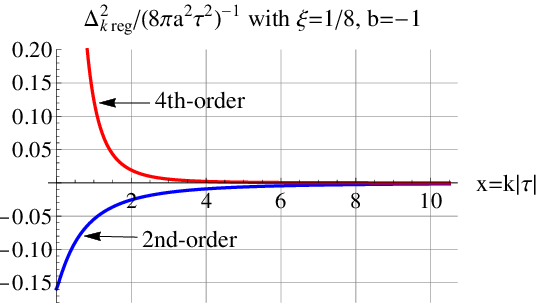}
\caption{
For  $b=-1$ and $\xi= \frac18  $,
   the 2nd-order regularized $\Delta_k^2$ is negative,
   the 4th-order  regularized $\Delta_k^2$ is positive and UV convergent.
      }
     \label{powerReb1Xi18new}
\end{figure}

\begin{figure}
\centering
\includegraphics[width=0.6\linewidth]{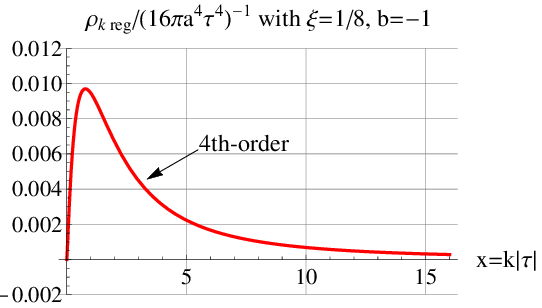}
\caption{
   For $b=-1$ and $\xi= \frac18 $,
   the  4th-order regularized $\rho_k$ is positive and UV convergent.
      }
     \label{4thrhob1Xi18new}
\end{figure}

Then, we consider  $b=-1$, de Sitter space.
We plot the  regularized power spectra
for $\xi=\frac{1}{8}$
in Fig.\ref{powerReb1Xi18new},
the 2nd-order regularized $\Delta_{k\, reg}^2$ is  negative,
the 4th-order regularized $\Delta_{k\, reg}^2$ is positive and UV convergent,
 but IR divergent.
As  shown in Fig.\ref{4thrhob1Xi18new},
the  4th-order regularized $\rho_k$ is positive and UV and IR convergent.
(Here $\frac{2a''' a'}{a^2} -\frac{a''\, ^2}{a^2}-\frac{4a'' a'\,^2}{a^3}=0$
for de Sitter space,
so that
the 2nd-order subtraction term  (\ref{rhoA22countx}) is equal to
the 4th-order  (\ref{rhoA42countx}),
and the 2nd-order regularized $\rho_k$ is equal to the 4th-order one.)

\begin{figure}
\centering
\includegraphics[width=0.6\linewidth]{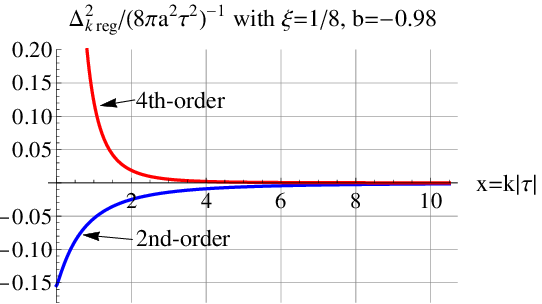}
\caption{
   For $b=-0.98$ and $\xi= 1/8 $,
    the  2nd-order regularized $\Delta_k^2$ is negative,
 the  4th-order  regularized $\Delta_k^2$  is positive and UV convergent.
      }
     \label{powerReb98Xi18new}
\end{figure}

\begin{figure}
\centering
\includegraphics[width=0.6\linewidth]{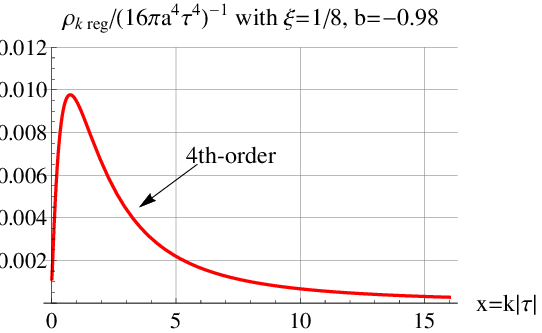}
\caption{
   For $b=-0.98$ and $\xi= 1/8 $,
 the    4th-order  regularized $\rho_k$  is positive and UV convergent.
      }
     \label{4thenergyReb98Xi18new}
\end{figure}

\begin{figure}
\centering
\includegraphics[width=0.6\linewidth]{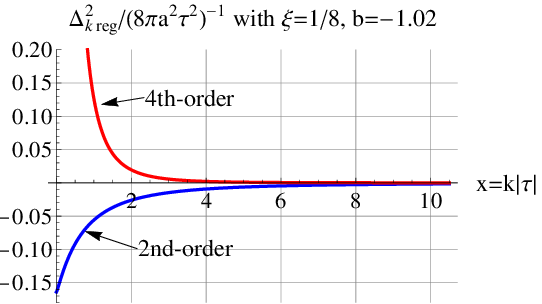}
\caption{
   For  $b=-1.02$ and $\xi=  1/8 $,
   The 2nd-order regularized $\Delta_k^2$ is negative,
   and the 4th-order regularized $\Delta_k^2$ is positive and UV convergent.
      }
     \label{powerReb102Xi18new}
\end{figure}

\begin{figure}
\centering
\includegraphics[width=0.6\linewidth]{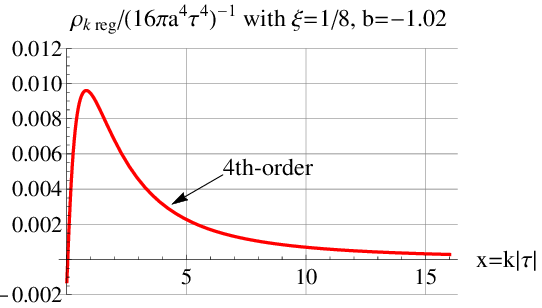}
\caption{
    For $b=-1.02$ and $\xi=\frac{1}{8}$,
    the 4th-order regularized $\rho_{k\, reg}$
    is positive at high $k$.}
     \label{4thrhb102Xi8new}
\end{figure}

Finally,
we consider the quasi de Sitter inflation model
with   $b= -0.98$ and  $\xi= \frac{1}{8}$.
As shown in Fig. \ref{powerReb98Xi18new},
the 2nd-order regularized $\Delta_k^2$ is negative,
the 4th-order regularized $\Delta_k^2$ is  positive and
UV convergent but IR divergent.
Fig.\ref{4thenergyReb98Xi18new} shows that
the 4th-order regularized $\rho_k$ is positive, IR finite and UV convergent.
For the  model with $b=-1.02$ and  $\xi=\frac{1}{8}$,
as  shown in Fig. \ref{powerReb102Xi18new},
the 2nd-order regularized $\Delta_k^2$ is negative;
the 4th-order regularized $\Delta_k^2$ is positive and UV convergent,
but IR divergent.
Fig.\ref{4thrhb102Xi8new} shows that
the 4th-order regularized $\rho_k$ is positive and UV convergent,
but  IR log divergent.

In the above when IR divergence appears,
the inside-horizon scheme can apply
as in Sect. \ref{section4},
and we do not repeat the details to save room.
Hence,   for general  $\xi$ and $b$,
 positive,  UV- and IR-convergent
power spectrum and spectral energy density
  can be achieved.

\section{Conclusion and Discussions}\label{section6}

We have studied  adiabatic regularization of
a  massless scalar field in general RW spacetimes
with a coupling $\xi$
(in a range $0 \leq \xi \leq  \frac{1}{6}$ for specific  in this paper),
and this  extends our previous study in de Sitter space \cite{ZhangYeWang2019}.
The analytical expressions of
the power spectrum,  the corresponding Green's function,
and the spectral stress tensor are given,
and all contain UV divergences.
Our goal is to find the appropriate schemes of regularization
that will   achieve nonnegative, UV- and IR-convergent
power spectrum and spectral energy density.
Adiabatic regularization
respects the covariant conservation of stress tensor to each order.
For the massless field,
the UV divergences are  generally removed
when the power spectrum is regularized to the 2nd-order,
and  respectively the stress tensor to the 4th order.
The nonnegativeness, however, is not always ensured
by the  conventional prescription,
and this is our main concern in this paper.
Through several examples,
we have found  that
there is no  regularization scheme of fixed-order
which   would work for all coupling  and  all RW spacetimes.
An adequate scheme   depends  upon the coupling $\xi $
and the expansion index  $b$.

Several   interesting cases are very simple.
For the   conformally-coupling  massless scalar field in Sect \ref{section3},
we  found that the regularized power spectrum and stress tensor are zero,
no trace anomaly exists in general  RW spacetimes.
The  regularization of the field with conformal symmetry
effectively amounts to the normal-ordering in the Minkowski spacetime.
This result is a generalization of
our previous work on de Sitter space \cite{ZhangYeWang2019}.
We have also explicitly identified  the mistakes
of Refs.\cite{Wald1978,DaviesFullingChristensenBunch1977}
on a massless field.
Most literature on the trace anomaly started with a massive field,
adopted the 4th-order regularization on the stress tensor,
and then took the massless limit.
As we showed in details in Ref.\cite{ZhangYeWang2019},
for a massive scalar field,
the 4th-order regularization is not appropriate,
because it will lead to an unphysical, negative spectral energy density,
which is a vital shortcoming unnoticed in the previous literature.
In fact, for a conformally-coupling  massive scalar field,
the 0th-order regularization \cite{ZhangYeWang2019} is the correct scheme,
as it not only removes all the UV divergences,
but also gives a positive spectral energy density,
and the resulting stress tensor is zero in the massless limit,
 agreeing  with the present paper.
Therefore, The trace anomaly  claimed in literature
is an artifact caused by the inadequate 4th-order regularization.

Another simple case is   minimally-coupling $\xi=0$ in Sect \ref{section4},
the regularized spectra  are zero
for   $b=0, \pm 1, 2$,
corresponding to  the Minkowski spacetime,  the de Sitter space,
the RD stage,  and  the matter-dominated stage.
In the  above simple cases,
we  also conducted carried out direct regularization of Green's functions in position space,
and found that  the regularized  Green's functions are zero as well,
confirming the zero spectra by adiabatic regularization.
In particular,  for the  RD stage,
the regularized spectra are also zero for any coupling $\xi$.
This is because  during the RD stage  the scalar curvature $R=0$
so that the wave equation (\ref{fieldequxi}) with arbitrary $\xi$
 is also conformal to those  in the Minkowski spacetime.

For the cases of general $\xi$ and $b$  in Sect \ref{section5},
 we  found that
the regularized spectra of the massless scalar field  can be negative
under the conventional   regularization.
To avoid the negative spectra,
we  performed higher order regularization to the pertinent spectrum.
Specifically,
if the  power spectrum  is negative under the 2nd-order regularization,
we calculate its 4th-order regularization,
and similarly if the spectral energy density  is negative
under the  4th-order regularization,
we calculate its 6th-order regularization.
In fact, the resulting higher-order regularization spectrum
will usually  become positive for the RW spacetimes commonly used in cosmology.
In some rarely-used RW spacetimes, the 6th-order regularized
spectral energy density may be still negative,
then we go to the 8th-order
which will eventually yield a positive spectral energy density.

A  massless field may  carry IR divergence  in the  unregularized spectra
as summarized in (\ref{spLowk1}) (\ref{enersmallk}) (\ref{presssmallk}).
Refs.\cite{FordParker1977,VilenkinFord1982}
present  studies regarding avoiding  the IR divergence in the  unregularized spectra.
In this paper we have analyzed
the IR divergences   caused by regularization.
In particular, IR divergence occurs when going to higher order regularization
 for  general $\xi$ and  $b$.
These new  IR divergences can be avoided by
the inside-horizon scheme of regularization,
as demonstrated by the examples in Fig.\ref{rhoInsideH}
and Fig.\ref{powerInsideH}.
The details of inside-horizon scheme were discussed in Ref. \cite{ZhangWangJCAP2018},
and we just  mention the following two related points.
First, the  scalar field considered in this paper is  linear,
 and its  $k$-modes are independent of each other,
unlike the nonlinear fields \cite{WangZhang2017}.
Under the inside-horizon scheme,
each regularized short wavelength mode respects
the covariant conservation to pertinent order,
the unregularized long wavelength modes also  respect  the covariant conservation.
Thus the total stress tensor  respects the covariant conservation
by  the inside-horizon scheme.
Second,  the long wavelength modes outside the horizon correspond to
the wave band of the observed CMB temperature anisotropies and polarization
\cite{WangZhangChen2016,ZhangWangJCAP2018}.
When the scalar field is used to model the cosmological perturbations,
the inside-horizon scheme reserves the spectra perturbations
of  the long wavelength band,
so that the observed primordial spectrum will not affected
by   regularization.

Through   these detailed  investigations,
we come to the conclusion:
for a coupling massless scalar field  in general RW spacetimes,
the nonnegative, UV- and IR-convergent power spectrum and spectral energy density
can be achieved by adiabatic regularization,
with the help of higher order scheme
and the inside-horizon scheme when necessary.

\

\textbf{Acknowledgements}

Y. Zhang is supported by
NSFC Grant No. 11421303, 11675165, 11633001,   11961131007.
 B. Wang is supported by CPSF Grant No. 2019M662168.
The authors thank A. Marciano for valuable discussions.

\appendix
\numberwithin{equation}{section}

\section{  High $k$ expansions of exact modes }\label{sectionA}

We list some asymptotic expressions following from
the analytical solution for a general RW spacetime
which are used in the context.

At high   $k $,  the  mode $v_k$ of (\ref{u})
 approaches to
\bl \label{uinfl}
v_{k } \simeq &     \frac{1}{\sqrt{2k}}e^{ix }
     \Big(1 + i\frac{(4\nu^2-1)}{8x} -\frac{(16\nu^4 -40\nu^2 +9)}{128 x^2}
     -i\frac{(64\nu^6 -560\nu^4 +1036\nu^2 -225)}{3072 x^3}  \nn \\
&¡¡+   \frac{(256\nu^8-5376\nu^6+31584 \nu^4 -51664 \nu^2 +11025 )}{98304 x^4} \nn \\
&  + i \frac{  \left(-1024 \nu ^{10}+42240 \nu ^8-561792 \nu ^6
     +2764960 \nu ^4-4228884 \nu ^2+893025\right)}{3932160 x^5} + ...
    \Big) ,  \nn \\
\el
where the first term corresponds to the positive-frequency mode
in Minkowski spacetime,
and other terms are due to expansion effects.
The squared mode at high $k$ is
\bl\label{vksq}
|v_k|^2 = \frac{\pi x}{4k} \big|  H^{(2)}_{\nu} ( x)   \big|^2
= & \frac{1}{2k} \Big(  1+ \frac{4\nu^2 -1}{8 x^2}
       +\frac{3(16\nu^4 -40\nu^2 +9)}{128 x^4} \nn \\
&       +\frac{5(2\nu-5)(2\nu-3)(2\nu-1)(2\nu+1)(2\nu+3)(2\nu+5)}{1024 x^6}
         + ...  \Big).   \nn   \\
\el
The time   derivatives   to the 4th adiabatic order are  given by
{\allowdisplaybreaks
\bl\label{vkprsquare}
|v_k'|^2 = & k \l(\frac{1}{2} - \frac{  ( 4 \nu^2 -1)}{16 x^2}
         - \frac{ (16\nu^4 - 104\nu^2 +25)}{256 x^4}
         -\frac{ (64\nu^6 -2096 \nu^4 +4876\nu^2 -1089)}{2048 x^6}
       + ... .  \r) \nn \\
\el
\bl\label{vkaExp}
|(\frac{v_k}{a})'|^2
= &
b^{-2}H^2\Big(
\frac{x^2}{2 k}
+\frac{8 b^2+1-4 \nu ^2}{16 k}
-\frac{16 \nu ^4
-\left(64 b^2+128 b+104\right) \nu ^2
+(16 b^2+32 b+25)}{256 k x^2}
\nn\\
&
+\frac{(216 b^2+864 b+1089)
-\left(960 b^2+3840 b+4876\right) \nu ^2
+\left(384 b^2+1536 b+2096\right) \nu ^4
-64 \nu ^6}{2048 k x^4}
\nn\\
&
+...\Big) .
\el
}
From these, one obtains
the high-$k$ expansions of $\rho_k$ and $p_k$
 in Eq.(\ref{rhokHighExp}) and (\ref{presskHighExp})
in the context.

At low $k$, the  mode $v_k$ of (\ref{u}) and the related squared modes are given by
\be \label{ulow}
v_{k }  \simeq   (\frac{x}{2})^{-\nu + \frac12}
  \frac{  \Gamma(\nu)}{\sqrt{2 \pi k}}
                 e^{i \frac{\pi}{2}(\nu - \frac12) } ,
\ee
\be\label{sqvkm}
|v_k|^2 \simeq    x^{-2\nu } |\tau| \frac{ 2^{2\nu-2} \Gamma(\nu)^2 }{ \pi },
\ee
\be\label{sqvkma}
| \frac{v_k}{a}|^2 \simeq
a_0^{-2}x^{-2\nu } |\tau|^{1-2b} \frac{ 2^{2\nu-2} \Gamma(\nu)^2 }{ \pi },
\ee
and the time derivatives are
\be
v_k' \simeq   \frac{|\tau|}{\tau}
    (\frac{x}{2})^{-\nu - \frac12}
   k^{1/2}
  \frac{  \Gamma(\nu)-2\Gamma(\nu +1)}{4 \sqrt{2 \pi }}
                 e^{i \frac{\pi}{2}(\nu - \frac12) } ,
\ee
\be
|v_k'|^2  \simeq    (\frac{x}{2})^{-2\nu - 1}
   k
  \frac{  (\Gamma(\nu)-2\Gamma(\nu +1))^2 }{32 \pi},
\ee
\be
(\frac{v_k}{a})'=
\frac{|\tau|}{\tau} a_0^{-1}
\left(\frac{x}{2}\right)^{-\frac{1}{2}-b-\nu}
k^{\frac{1}{2}+b}
\frac{(1-2 b) \Gamma (\nu )-2 \Gamma (\nu +1) }{  2^{2+b}  \sqrt{2 \pi }}
e^{ i \frac{\pi}{2} (\nu -\frac{1}{2})}
   ,
\ee
\be \label{lwkvpsq}
|(\frac{v_k}{a})'|^2
 =
a_0^{-2} x^{-2 \nu }|\tau|^{-1-2 b}
\frac{ 2^{2 \nu } \Big[(1-2 b) \Gamma (\nu )-2 \Gamma (\nu +1)\Big]^2}{16 \pi }.
\ee
From these follow the  low-$k$  expansions  of $\rho_k$ and $p_k$
in  Eq.(\ref{enersmallk}) and  Eq.(\ref{presssmallk}).

\section{ The 0th-, 2nd-, and 4th-order adiabatic subtraction terms}
\label{sectionB}

The WKB  approximate  solution \cite{Chakraborty1973,ParkerFulling1974,
FullingParkerHu1974,Bunch1980,AndersonParker1987,BirrellDavies1982}
 of  the massless scalar field equation  (\ref{equvk})
 is   written as the following
\be\label{vn}
v_k^{(n)}(\tau)
  = (2W_k(\tau))^{-1/2}   \exp \Big[  -i \int^{\tau} W_k(\tau')d\tau' \Big],
\ee
where the effective frequency is
\be\label{equWk}
  W_k(\tau)     = \Big[  k^2  + (\xi -\frac16)a^2R
-\frac12 \l( \frac{ W_k  '' }{ W_k}
- \frac32 \big( \frac{W_k  '}{W_k} \big)^2 \r)  \Big]^{1/2}.
\ee
The WKB solution of $W_k$ is obtained by iteratively solving (\ref{equWk})
to a desired adiabatic  order.
Take  the 0th-order  \cite{Bunch1980},
\be
W_k^{(0)}=  k ,
\ee
and the 0th-order adiabatic  mode
\be\label{vkWKB0}
v_k^{(0)}(\tau)
  =(2 k)^{-1/2}
    e^{  -i \int^{\tau}k d\tau'} .
\ee
The 0th-order quantities that appear in the 0th-order subtraction terms are
\be\label{ModvkWKB0}
|v_k^{(0)}|^2=\frac{1}{2W_k^{(0)}}=\frac{1}{2k},
\ee
\bl
|v_k^{(0)\prime}|^2
=&
\frac{1}{2}\l(
\frac{(W_k^{(0)\prime})^2}{4(W_k^{(0)})^{3}}+W_k^{(0)} \r)
   = \frac{k}{2} ,
\el
\bl
   v_k^{(0)\prime}v_k^{(0)*}+v_k^{(0)}v_k^{(0)*\,\prime}=0 ,
\el
\bl
|(\frac{v^{(0)}_k}{a})' |^2    = &
 \frac{1}{a^2} \Big( |v^{(0)}_k \, ' |^2
 - \frac{a'}{a} (v_k ^{(0)}\, ' v_k^{(2)}\, ^*  + v_k^{(0)} v_k^{(2)}\, ^*\, ')
   +(\frac{a'}{a})^2 |v^{(0)}_k|^2     \Big)
 \nn\\
 = &
 \frac{1}{a^2} \Big( \frac{k}{2}
   +(\frac{a'}{a})^2 \frac{1}{2k}  \Big) .
\el
These 0th-order subtraction terms are independent of $\xi$.
The 2nd-order adiabatic  mode
\be\label{vkWKB2}
v_k^{(2)}(\tau)
  =(2W_k^{(2)}(\tau))^{-1/2}
    e^{  -i \int^{\tau} W_k^{(2)}(\tau')d\tau'} ,
\ee
The  2nd-order effective frequency is given by first iteration of (\ref{equWk})
\be \label{W2expression}
W_k^{(2)}= \l[ k^2   + (\xi -\frac16)a^2R
-\frac12 \l( \frac{ W^{(0)}_k\,  '' }{ W^{(0)} _k }
- \frac32 \big( \frac{W^{(0)}_k\,  '}{W^{(0)}_k} \big)^2 \r) \r]^{\frac12}
  \, .
\ee
Keeping only two time derivatives of $a(\tau)$
gives
\be\label{Wk2nd}
W_k^{(2)}=\sqrt{k^2+6 (\xi -\frac16) \frac{a''}{a}}
 \simeq    k
 + 3  (\xi-\frac16)\frac{1}{k } \frac{a''}{a},
\ee
\be\label{W2xi}
(W_k^{(2)})^{-1}
 \simeq   \frac{1}{k}
       - 3  (\xi-\frac16)  \frac{a''/a}{k^{3}}
    .
\ee
The 2nd-order subtraction term for the power spectrum is
\be\label{v2subsq}
|v^{(2)}_k|^2   =\frac{1}{2 W_k^{(2)}}
=\frac{1}{2k}
       - \frac{3}{2}  (\xi-\frac16)  \frac{a''/a}{k^{3}}.
\ee
The  2nd-order subtraction term for $\rho_k$ and $p_k$
also involves the following terms
\be
|v^{(2)}_k \, ' |^2  = \frac12
      \Big(  \frac{ (W_k^{(2)\, '})^2 }{4 (W_k^{(2)})^{3} } +  W_k^{(2)}  \Big)
   =\frac{k}{2}   +(\xi-\frac16)\frac{3    a''}{2 k a } ,
\ee
\bl
v_k ^{(2)}\, ' v_k^{(2)}\, ^*  + v_k^{(2)} v_k^{(2)}\, ^*\, '
& =  -   \frac{ W_k^{(2)\, '} }{2 ( W_k^{(2)})^2 }
  \simeq 0  ,
\el
\bl\label{energyk}
|(\frac{v^{(2)}_k}{a})' |^2
= &  \frac{1}{a^2} \Big( |v^{(2)}_k \, ' |^2
 - \frac{a'}{a} (v_k ^{(2)}\, ' v_k^{(2)}\, ^*  + v_k^{(2)} v_k^{(2)}\, ^*\, ')
   +(\frac{a'}{a})^2 |v^{(2)}_k|^2     \Big)  \nn \\
=&   \frac{k}{2a^2}+\frac{1}{2k a^2}(\frac{a'}{a})^2
   +(\xi-\frac16)\frac{3}{2k } \frac{a''}{a^3}   .
\el
These 2nd-order subtraction terms  dependent on $\xi$.
A very important property of a massless scalar field is that
$W_k^{(2)} =W_k^{(0)}=k$ for the conformal coupling $\xi=\frac16$.
Actually, $W_k^{(n)} =W_k^{(0)}=k$ holds for an arbitrary $n$th-order,
as is implied by the iteration formula (\ref{equWk}).

The 4th-order adiabatic  mode is defined by
\be\label{vkWKB4}
v_k^{(4)}(\tau)
  =(2W_k^{(4)}(\tau))^{-1/2}
    e^{  -i \int^{\tau} W_k^{(4)}(\tau')d\tau'} ,
\ee
the 4th-order effective frequency is given   by iteration
\be\label{Wk4}
W_k^{(4)}= \l[ k^2  + (\xi -\frac16)a^2R
-\frac12 \l( \frac{ W^{(2)}_k\,  '' }{ W^{(2)} _k }
- \frac32 \big( \frac{W^{(2)}_k\,  '}{W^{(2)}_k} \big)^2 \r) \r]^{\frac12} .
\ee
Keeping up to four  time derivatives,
one obtains
\bl\label{W4s}
W_k^{(4)} =& k +   \frac{3(\xi-\frac16)   }{k   } \frac{a''}{a}
          -\frac92  \Big(\xi-\frac16 \Big)^2 \frac{a''\, ^2}{a^2 k^3}   \nn \\
& - (\xi -\frac{1}{6})
   \frac{3}{4 k^3} \Big( \frac{ a''''}{a}-\frac{ a''\, ^2}{a ^2}
     -\frac{2 a'''  a' }{a ^2}+\frac{2 a'\, ^2 a''}{ a^3} \Big)
     ,\nn \\
\el
 and
\bl \label{invW4s}
(W_k^{(4)})^{-1} =&
   \frac{1}{k } -3 (\xi -\frac{1}{6} )  \frac{1}{k ^3} \frac{a''}{a}
   +  (\xi -\frac{1}{6}) \frac{3}{4 k^5} (\frac{a'''' }{a} -\frac{a''\, ^2}{a^2}
   + 2 \frac{a'\, ^2 a''}{a^3}  -2 \frac{a'''  a'}{a^2})    \nn \\
& + (\xi -\frac{1}{6})^2 \, \frac{27}{2 k^5} \frac{a''\,^2}{a^2}  .
\el
Then
\be\label{v2sub}
|v^{(4)}_k|^2   =  (2 W_k^{(4)})^{-1} ,
\ee
\bl
|v^{(4)}_k \, ' |^2
=& \frac{k}{2}   +(\xi-\frac16)\frac{3 a''}{2 k a}
 - (\xi -\frac{1}{6} )^2\frac{9}{4 k^3} \frac{a''^2}{a^2}   - (\xi -\frac{1}{6} )
        \frac{3}{8 k ^3}  (\frac{  a''''}{a}
     -\frac{a''^2}{a^2} -\frac{2 a''' a'}{a^2} +\frac{2 a'^2 a''}{a^3} ) , \nn \\
\el
\bl
(v_k ^{(4)}\, ' v_k^{(4)}\, ^*  + v_k^{(4)} v_k^{(4)}\, ^*\, ')
& =   -   \frac{ W_k^{(4)\, '} }{2 ( W_k^{(4)})^2 }
  =   -\frac{3}{2} (\xi -\frac{1}{6})
    ( \frac{ a'''}{  k^3 a}-\frac{a'a''}{ k^3 a^2 }),
\el
putting together yields
{\allowdisplaybreaks
\bl
|(\frac{v^{(4)}_k}{a})' |^2 =& \frac{1}{a^2} \Bigg[|v_k^{(4)'}|^2
 -\frac{a'}{a}(v_k^{(4)'}v_k^{(4)*}+v_k^{(4)}v_k^{(4)*'})
 +(\frac{a'}{a})^2|v_k^{(4)}|^2 \Bigg] \nn \\
=  &    \frac{1}{a^2} \Bigg[\frac{k }{2}
+\frac{1}{2  k } \frac{a'^2}{ a^2}
+(\xi-\frac16)\frac{3  a''}{2 k  a }
-(\xi-\frac16)^2 \frac{9}{4 k^3} \frac{a''^2}{a^2} \nn \\
& -(\xi-\frac16)  \frac{3}{8 k^3}(\frac{ a''''}{a} -\frac{a''^2}{a^2}
     -\frac{6 a' a'''}{ a^2} +\frac{10 a'^2 a''}{ a^3})  \Bigg].
\el
}
These 4th-order subtraction terms  also dependent on $\xi$.
The portions up to two time derivatives of the above
reduce to the 2nd-order results.

Similarly, the 6th-order adiabatic  mode
\be\label{vkWKB6}
v_k^{(6)}(\tau)
  =(2W_k^{(4)}(\tau))^{-1/2}
    e^{  -i \int^{\tau} W_k^{(4)}(\tau')d\tau'} .
\ee
The  6th-order effective frequency is derived  by iteration
\be\label{Wk6}
W_k^{(6)}= \l[ k^2  + (\xi -\frac16)a^2R
-\frac12 \l( \frac{ W^{(4)}_k\,  '' }{ W^{(4)} _k }
- \frac32 \big( \frac{W^{(4)}_k\,  '}{W^{(4)}_k} \big)^2 \r) \r]^{\frac12} .
\ee
Keeping up to six time derivatives,  one obtains
\bl\label{W6s}
W_k^{(6)}
= &
k
+\frac{3 (\xi-\frac{1}{6})}{k}\frac{ a''}{ a}
-\frac{9 }{2}(\xi-\frac{1}{6}) ^2 \frac{a''^2}{ a^2 k^3}
-(\xi-\frac{1}{6})\frac{3}{4k^3}\left(\frac{ a''''}{a}
    -\frac{a''^2}{a^2}
    -\frac{2 a''' a'}{ a^2}
    +\frac{2 a'^2 a''}{a^3}\right)
\nn\\
&
+\frac{ 27 }{2}(\xi-\frac{1}{6}) ^3\frac{a''^3}{ a^3 k^5}
+\frac{(\xi-\frac{1}{6}) ^2}{k^5}\left(\frac{45 a'''^2}{8 a^2}
    -\frac{27 a''^3}{4 a^3}
    +\frac{27 a'''' a''}{4 a^2}
    +\frac{153 a'^2 a''^2}{8 a^4}
    -\frac{99 a''' a' a''}{4 a^3}\right)
\nn\\
&
+\frac{(\xi-\frac{1}{6})  }{k^5}\Big(\frac{3 a''''''}{16 a}
    -\frac{3 a'''^2}{4 a^2}
    +\frac{9 a''^3}{8 a^3}
    -\frac{3 a''''' a'}{4 a^2}
    -\frac{21 a'''' a''}{16 a^2}
    +\frac{9 a'''' a'^2}{4 a^3}
    -\frac{9 a''' a'^3}{2 a^4}
\nn\\
&
    +\frac{9 a'^4 a''}{2 a^5}
    -\frac{27 a'^2 a''^2}{4 a^4}
    +\frac{6 a''' a' a''}{a^3}\Big),
\el
 and
\bl \label{invW6s}
(W_k^{(6)})^{-1} =&
\frac{1}{k}
-\frac{3(\xi-\frac{1}{6})}{k^3 }\frac{ a''}{a}
+\frac{3(\xi-\frac{1}{6})}{4 k^5}\left(
\frac{ a''''}{ a}
-\frac{ a''^2}{ a^2}
+\frac{2 a'^2 a''} {a^3}
-\frac{2a''' a'}{ a^2}\right)
+\frac{27(\xi-\frac{1}{6}) ^2 }{2 k^5 }\frac{a''^2}{a^2}
\nn\\
&
-\frac{135 (\xi-\frac{1}{6}) ^3 a''^3}{2 k^7 a^3}
+\frac{(\xi-\frac{1}{6}) ^2}{k^7}
\Big(-\frac{45 a'''^2}{8 a^2}+\frac{45 a''^3}{4 a^3}
-\frac{45 a'''' a''}{4 a^2}-\frac{225 a'^2 a''^2}{8 a^4}
\nn\\
&
+\frac{135 a''' a' a''}{4 a^3}\Big)
+\frac{(\xi-\frac{1}{6})  }{k^7}
\Big(-\frac{3 a''''''}{16 a}+\frac{3 a'''^2}{4 a^2}
-\frac{9 a''^3}{8 a^3}+\frac{3 a''''' a'}{4 a^2}+\frac{21 a'''' a''}{16 a^2}
\nn\\
&
-\frac{9 a'''' a'^2}{4 a^3}+\frac{9 a''' a'^3}{2 a^4}
-\frac{9 a'^4 a''}{2 a^5}+\frac{27 a'^2 a''^2}{4 a^4}
-\frac{6 a''' a' a''}{a^3}\Big) .
\el
Then
\be\label{v6sub}
|v^{(6)}_k|^2   =  (2 W_k^{(6)})^{-1} ,
\ee
\bl
|v^{(6)}_k \, ' |^2  =& \frac12
      \Big(  \frac{ (W_k^{(6)\, '})^2 }{4 (W_k^{(6)})^{3} } +  W_k^{(6)}  \Big)
           \nn \\
=&
\frac{k}{2}
+\frac{3(\xi-\frac{1}{6}) a''}{2 k a}
-\frac{9(\xi-\frac{1}{6}) ^2 a''^2}{4 k^3 a^2}
-\frac{3(\xi-\frac{1}{6})}{8 k^3}\left(\frac{ a''''}{ a}
-\frac{ a''^2}{ a^2}
-\frac{2 a''' a'}{ a^2}
+\frac{2 a'^2 a''}{ a^3}\right)
\nn\\
&
+\frac{27(\xi-\frac{1}{6}) ^3  a''^3}{4 k^5 a^3}
+\frac{(\xi-\frac{1}{6}) ^2}{k^5}\Big(\frac{63 a'''^2}{16 a^2}
-\frac{27 a''^3}{8 a^3}
+\frac{27 a'''' a''}{8 a^2}
+\frac{171 a'^2 a''^2}{16 a^4}
-\frac{117 a''' a' a''}{8 a^3}\Big)
\nn\\
&
+\frac{(\xi-\frac{1}{6})}{k^5}\Big(\frac{3 a''''''}{32 a}
-\frac{3 a'''^2}{8 a^2}
+\frac{9 a''^3}{16 a^3}
-\frac{3 a''''' a'}{8 a^2}
-\frac{21 a'''' a''}{32 a^2}
+\frac{9 a'''' a'^2}{8 a^3}
-\frac{9 a''' a'^3}{4 a^4}
\nn\\
&
+\frac{9 a'^4 a''}{4 a^5}
-\frac{27 a'^2 a''^2}{8 a^4}
+\frac{3 a''' a' a''}{a^3}\Big)
 ,
\el
\bl
(v_k ^{(6)}\, ' v_k^{(6)}\, ^*  + v_k^{(6)} v_k^{(6)}\, ^*\, ')
   =   &   -   \frac{ W_k^{(6)\, '} }{2 ( W_k^{(6)})^2 }   \nn\\
  =  &  -\frac{3(\xi-\frac{1}{6})}{2 k^3}
  \left(\frac{a'''}{ a}-\frac{a' a''}{a^2} \right)
+\frac{(\xi-\frac{1}{6}) ^2}{k^5}\left(\frac{27 a''' a''}{2 a^2}
   -\frac{27 a' a''^2}{2 a^3}\right)   \nn\\
&   +\frac{(\xi-\frac{1}{6})}{k^5} \left(\frac{3 a'''''}{8 a}
-\frac{9 a'''' a'}{8 a^2}
-\frac{3 a''' a''}{2 a^2}
+\frac{9 a''' a'^2}{4 a^3}
-\frac{9 a'^3 a''}{4 a^4}
+\frac{9 a' a''^2}{4 a^3}\right),
\el
putting together yields
\bl
|(\frac{v^{(6)}_k}{a})' |^2 =& \frac{1}{a^2} \Bigg[|v_k^{(6)'}|^2
 -\frac{a'}{a}(v_k^{(6)'}v_k^{(6)*}+v_k^{(6)}v_k^{(6)*'})
 +(\frac{a'}{a})^2|v_k^{(6)}|^2 \Bigg]
\nn \\
=&
\frac{1}{a^2}
\Bigg[
\frac{k}{2 }
+\frac{a'^2}{2 k a^2}
+\frac{3 (\xi-\frac{1}{6})  a''}{2 k a}
-\frac{9 (\xi-\frac{1}{6}) ^2 a''^2}{4 k^3 a^2}
-\frac{3(\xi-\frac{1}{6})}{8k^3}\Big(\frac{ a''''}{ a}
-\frac{ a''^2}{ a^2}
-\frac{6 a''' a'}{a^2}
\nn\\
&
+\frac{10 a'^2 a''}{ a^3}\Big)
+\frac{27(\xi-\frac{1}{6}) ^3  a''^3}{4 k^5 a^3}
+\frac{(\xi-\frac{1}{6}) ^2}{k^5}\Big(\frac{63 a'''^2}{16 a^2}
-\frac{27 a''^3}{8 a^3}
+\frac{27 a'''' a''}{8 a^2}
+\frac{495 a'^2 a''^2}{16 a^4}
\nn\\
&
-\frac{225 a''' a' a''}{8 a^3}\Big)
+\frac{(\xi-\frac{1}{6})}{k^5}\Big(\frac{3 a''''''}{32 a}
-\frac{3 a'''^2}{8 a^2}+\frac{9 a''^3}{16 a^3}
-\frac{3 a''''' a'}{4 a^2}
-\frac{21 a'''' a''}{32 a^2}
+\frac{21 a'''' a'^2}{8 a^3}
\nn\\
&
-\frac{21 a''' a'^3}{4 a^4}
+\frac{21 a'^4 a''}{4 a^5}
-\frac{6 a'^2 a''^2}{a^4}
+\frac{9 a''' a' a''}{2 a^3}\Big)
\Bigg]
.
\el
The portions up to four  time derivatives of the above
reduce to the 4th-order results.

The 0th-order subtraction term for spectral energy density and pressure
\bl \label{rhoA0}
\rho_{k\, A0}  =& \frac{ k^3}{4\pi^2 a^4}
 \Big[ |v^{(0)} _k\, '|^2 + k^2  |v^{(0)}_k|^2  \nn \\
&   + (6\xi-1) \Big(
   \frac{a'}{a} (v^{(0)}_k\, '  v^{(0)}_k\, ^* + v_k^{(0)} v^{(0)}_k\, ^*\, ' )
    -  \frac{a'\, ^2}{a^2}  |v^{(0)}_k|^2  \Big) \Big]
 =   \frac{ k^4}{4\pi^2 a^4} ,
\el
\bl \label{pkA0}
p_{k\, A 0}
=& \frac{k^3}{4 \pi^2 a^4}
  \Bigg[   \frac13 |v^{(0)} _k\, '|^2 + \frac13 k^2  |v^{(0)}_k|^2
      + 2(\xi-\frac16)\Big( -2 |v^{(0)} _k\, '|^2
       + 3 \frac{a'}{a} (v^{(0)}_k\, '  v^{(0)}_ k\, ^* + v_k^{(0)} v^{(0)}_k\, ^*\, ')
       \nn \\
&  - 3(\frac{a'}{a})^2 |v_k^{(0)}|^2
    +  2 k^2  |v_k^{(0)}|^2
    + 12 \xi  \frac{a''}{a} |v_k^{(0)}|^2  \Big)\Bigg]
  =  \frac{k^4}{12 \pi^2 a^4} .
\el
which are independent of $\xi$.

The  2nd-order subtraction term for
the spectral energy density and pressure
\bl   \label{rhoA22countx}
\rho_{k\,A 2}
= & \frac{k^4}{4\pi^2 a^4}
\Big[  1
   - ( \xi-\frac16) \frac{3}{k^2} \frac{a'\,^2}{a^2}  \Big] \nn \\
= & \frac{1}{4\pi^2 a^4\tau^4}\Big[ x^4 -\frac{(6 \xi-1)b^2 x^2}{2 }\Big],
\el
\bl \label{pA2counxt}
p_{k\, A 2} =  &  \frac{k^4}{12 \pi^2 a^4}
\Big[1 + (\xi-\frac16) \frac{1}{k^2} \Big(6\frac{a''}{a}- 9\frac{a'\,^2}{a^2}
           \Big) \Big] \nn \\
  = &  \frac{1}{4 \pi^2 a^4\tau^4}
      \Big[ \frac{x^4}{3}-\frac{(6 \xi -1)b (b+2) x^2}{6} \Big]  ,
\el
which depend  on $\xi$.
It is important that
the derivatives are   kept only up to second order in these 2nd-order
subtraction terms,
and this will ensure the covariant conservation to the 2nd adiabatic order.

The  4th-order subtraction term for
the spectral energy density and pressure
\bl   \label{rhoA42countx}
\rho_{k\,A 4}
=& \frac{k^4}{4\pi^2 a^4}
\Big[  1
  - ( \xi-\frac16)   \frac{3}{k^2} \frac{a'\,^2}{a^2}
  - (\xi-\frac16)^2  \frac{9}{2 k^4}
    \Big( \frac{2a'''  a'}{a^2} -\frac{a'' \, ^2}{a^2}-\frac{4a''  a'\,^2}{a^3}\Big)
     \Big] \nn \\
= & \frac{1}{4\pi^2 a^4\tau^4}\Bigg[
    x^4 -\frac{(6 \xi-1)b^2 x^2}{2 }+\frac{3 (6 \xi -1)^2(b-1)b^2 (b+1) }{8}\Bigg],
\el
\bl\label{pA4counxt}
p_{k\, A 4} =&   \frac{k^4}{12 \pi^2 a^4}
\Bigg[1 + (\xi-\frac16) \frac{1}{k^2} \Big(6\frac{a''}{a}- 9\frac{a'\,^2}{a^2} \Big)
                \nn \\
& + (\xi-\frac16)^2 \frac{9}{2 k^4}
    \Big( \frac{2a''''  }{a}-\frac{10  a''' a'}{a^2}
     -\frac{5 a'' \, ^2}{a^2}   +\frac{16a''  a'\,^2}{a^3} \Big)\Bigg]  \nn \\
= &\frac{1}{4 \pi^2 a^4\tau^4} \Bigg[
  \frac{x^4}{3}-\frac{(6 \xi -1)b (b+2) x^2}{6}
  +\frac{ (6 \xi -1)^2(b-1)b(b+1) (b+4)}{8}
  \Bigg] .
\el

The 6th-order subtraction term for spectral energy density and pressure
\bl   \label{rhoA6countx}
\rho_{k\,A 6}
=& \frac{k^4}{4\pi^2 a^4}
\Big[
1
-\frac{3 (\xi-\frac{1}{6})  a'^2}{k^2 a^2}
-\frac{9(\xi-\frac{1}{6}) ^2}{2 k^4}\left(
\frac{2 a''' a'}{a^2}
-\frac{ a''^2}{ a^2}
-\frac{4 a'^2 a''}{a^3}\right)
\nn\\
&
+\frac{(\xi-\frac{1}{6}) ^3}{k^6}\left(
-\frac{27 a''^3}{a^3}
-\frac{243 a'^2 a''^2}{2 a^4}
+\frac{81 a''' a' a''}{a^3}\right)
+\frac{(\xi-\frac{1}{6}) ^2}{k^6}\Big(\frac{9 a'''^2}{8 a^2}
+\frac{9 a''^3}{4 a^3}+\frac{9 a''''' a'}{4 a^2}
\nn\\
&
-\frac{9 a'''' a''}{4 a^2}
-\frac{9 a'''' a'^2}{a^3}
+\frac{18 a''' a'^3}{a^4}
-\frac{18 a'^4 a''}{a^5}
+\frac{99 a'^2 a''^2}{8 a^4}
-\frac{27 a''' a' a''}{4 a^3}\Big) \Big] \nn \\
= &\frac{1}{4\pi^2 a^4\tau^4}\Bigg[
x^4 -\frac{(6 \xi-1)b^2 x^2}{2 }+\frac{3 (6 \xi -1)^2(b-1)b^2 (b+1) }{8}
\nn\\
& +\frac{5(6 \xi -1)^2 (b-1) b^2 (b+2)\l[(1- 6\xi)  (b^2-b)-2\r] }{16 x^2} \Bigg]
\el
which is the first four terms of  $\rho_{k}$ in (\ref{rhokHighExp}),
{\allowdisplaybreaks
\bl\label{pA6counxt}
p_{k\, A 6} =&
\frac{k^4}{12 \pi^2 a^4}
\Bigg[1
+\frac{(\xi-\frac{1}{6})}{k^2}\left(\frac{6 a''}{a}
-\frac{9 a'^2}{a^2}\right)
+\frac{9(\xi-\frac{1}{6}) ^2}{2k^4}\left(\frac{ 2a''''}{a}
-\frac{10 a''' a'}{a^2}
-\frac{5 a''^2}{a^2}
+\frac{16 a'^2 a''}{a^3}\right)
\nn\\
&
+\frac{(\xi-\frac{1}{6}) ^3}{k^6}\Big(-\frac{81 a'''^2}{a^2}
+\frac{135 a''^3}{a^3}
-\frac{81 a'''' a''}{a^2}
-\frac{1215 a'^2 a''^2}{2 a^4}
+\frac{567 a''' a' a''}{a^3}\Big)
\nn\\
&
+\frac{(\xi-\frac{1}{6}) ^2}{k^6}\Big(-\frac{9 a''''''}{4 a}
+\frac{81 a'''^2}{8 a^2}
-\frac{63 a''^3}{4 a^3}
+\frac{63 a''''' a'}{4 a^2}
+\frac{18 a'''' a''}{a^2}
-\frac{54 a'''' a'^2}{a^3}
\nn\\
&
+\frac{108 a''' a'^3}{a^4}
-\frac{108 a'^4 a''}{a^5}
+\frac{1071 a'^2 a''^2}{8 a^4}
-\frac{423 a''' a' a''}{4 a^3}\Big)\Bigg]  \nn \\
= & \frac{1}{4 \pi^2 a^4\tau^4} \Bigg[
 \frac{x^4}{3} -\frac{(6 \xi -1)b (b+2) x^2}{6}
 +\frac{ (6 \xi -1)^2(b-1)b(b+1) (b+4)}{8}
\nn\\
& + \frac{5(6 \xi -1)^2(b-1) b(b+2) (b+6)  \l[(1- 6\xi)  (b^2-b)-2\r]}{48 x^2}\Bigg]
\el
}
which is the first four terms of $p_k$ in (\ref{presskHighExp}).

As we have mentioned earlier,
for the conformally-coupling  $\xi=\frac16$,
one has
\be\label{W246}
(W_k^{(0)})^{-1}=(W_k^{(2)})^{-1}=(W_k^{(4)})^{-1}=(W_k^{(6)})^{-1}=\frac{1}{k},
\ee
\be\label{v246}
|v^{(0)}_k|^2 =|v^{(2)}_k|^2 =|v^{(4)}_k|^2 =|v^{(6)}_k|^2= \frac{1}{2k},
\ee
\be\label{rhoA246}
\rho_{k\,A 0}=\rho_{k\,A 2}=\rho_{k\,A 4}=\rho_{k\,A 6},
\ee
\be\label{pressureA246}
p_{k\,A 0}=p_{k\,A 2}=p_{k\,A 4}=p_{k\,A 6},
\ee
so that the 0th-, 2nd-,   4th- and 6th-order subtraction terms are all equal.
Inspection of iteration (\ref{equWk})
tells that, for  $\xi=\frac16$,
the subtraction terms of any order
are the same as (\ref{W246})--(\ref{pressureA246}).

Now we show  that the four-divergence of the subtraction terms
of the stress tensor is zero at each adiabatic order
\be\label{TA0}
\left\langle T^{\mu \nu}\,_{ ; \nu}\right\rangle_{A \, n}=0.
\ee
Time derivative of Eq.(\ref{rhoA6countx}) gives
\bl
\rho_{k A 6}^{\prime}= &
\frac{1}{16 \pi ^2 k^2 a^{10}}a' \Bigg [648 (\xi-\frac{1}{6}) ^2 a'^4 a''
-3 (\xi-\frac{1}{6})  a^4 \left(-3 (\xi-\frac{1}{6})  a''''''
+12 k^2 (\xi-\frac{1}{6})  a''''+8 k^4 a''\right) \nn \\
& +36 (\xi-\frac{1}{6}) ^2 a a'^2
\left((108 (\xi-\frac{1}{6}) -19) a''^2-18 a''' a'\right) \nn \\
& -36 (\xi-\frac{1}{6}) ^2 a^2 \left((6 (\xi-\frac{1}{6}) -1) a''^3+a'^2
\left(14 k^2 a''-9 a''''\right)+2 (45 (\xi-\frac{1}{6}) -7) a''' a' a''\right) \nn \\
& +9 (\xi-\frac{1}{6})  a^3 \bigg(8 k^4 a'^2+2 (\xi-\frac{1}{6})
  \left(16 k^2 a'''-5 a'''''\right) a' \nn \\
&+(\xi-\frac{1}{6})
  \Big(6 (6 (\xi-\frac{1}{6}) -1) a'''^2+4 k^2 a''^2
  +(36 (\xi-\frac{1}{6}) -5) a'''' a'' \Big) \bigg)
  -16 k^6 a^5 \Bigg] ,
\el
and combining  (\ref{rhoA6countx}) and (\ref{pA6counxt}) gives
\be
3 \frac{a^{\prime}}{a}\left(\rho_{k \, A 6}+p_{k \, A 6}\right)
=- \rho_{k A 6}^{\prime},
\ee
so that
the four-divergence of the 6th-order subtraction terms is zero
\be \label{6cconserv}
\rho_{k A 6}^{\prime}+3 \frac{a^{\prime}}{a}
\left(\rho_{k \, A 6}+p_{k \, A 6}\right)=0.
\ee
Similarly,
it is  checked that
\be\label{cconserv}
\rho_{k \, A \, n}^{\prime}
+3 \frac{a^{\prime}}{a}\left(\rho_{k \, A \, n}+p_{k \, A \, n}\right)=0
\ee
is valid for $n=0,2,4$.
Hence,
the covariant conservation
\be \label{cconservres}
\left\langle T^{\mu \nu(n)} \,_{; \nu}\right\rangle_{r e g}
  =\left\langle T^{\mu \nu}\,_{ ; \nu}\right\rangle
  -\left\langle T^{\mu \nu}\,_{ ; \nu}\right\rangle_{A \, n}= 0
\ee
is respected by the regularized stress tensor
at each order we need in this paper.

\end{document}